\documentclass[11pt]{article}
\usepackage{amsmath,amssymb,amsthm}
\usepackage{graphics,epsfig}
\usepackage{hyperref}
\usepackage{natbib}
\usepackage{color}
\usepackage{graphicx}
\usepackage{caption}
\usepackage{subcaption}
\usepackage{float}
\usepackage{dsfont}
\usepackage{mathrsfs}
\usepackage{multirow}
\usepackage{comment}
\usepackage{setspace}
\usepackage{bbm}
\usepackage{bm}
\usepackage{amsfonts}
\usepackage{xcolor}
\usepackage{pslatex}
\usepackage{setspace}
\usepackage{statrep}  % for SAS
\def \E {\mathbb{E}}
\def \P {\mathbb{P}}

\newtheorem{example}{\bf Example}
\newtheorem{exm}[example]{\bf Example}

\newtheorem{definition}{\bf Definition}

	\newtheorem{remark}{\bf Remark}
	\newtheorem{rmk}[remark]{\bf Remark}


\begin{document}
	
	\title{\bfseries A Bayesian Discrete Framework for Enhancing Decision-Making Processes in Clinical Trial Designs and Evaluations}
	
	\author{
		Paramahansa Pramanik \footnote{e-mail: {\small\texttt{parampramanik@outlook.com}} }\; \footnote{Department of Mathematics and Statistics, University of South Alabama, 411 North University Boulevard, Mobile, AL 36688, USA.} \; \; Arnab Kumar Maity \footnote{email:{\small\texttt{arnab.maity@boehringer-ingelheim.com}}}\; \footnote{Boehringer Ingelheim Pharmaceuticals, Inc., 900 Ridgebury Road, Ridgefield, CT 06877, USA.}\; \; Anjan Mandal\footnote{email:{\small\texttt{9.anjan@gmail.com}}}\;\footnote{Department of Mathematical Sciences, University of Nevada, Las Vegas, 4505 S. Maryland Pkwy.
Las Vegas, NV 89154, USA.}\; \;Haley Kate Robinson \footnote{email:{\small\texttt{hkr2322@jagmail.southalabama.edu}}}\;\footnote{Department of Biomedical Sciences,
University of South Alabama,
Mobile, Alabama 36688, USA.}}
	
	\date{\today}
	\maketitle
	
\subparagraph{Abstract.}
This study examines the application of Bayesian approach in the context of clinical trials, emphasizing their increasing importance in contemporary biomedical research. While conventional frequentist approach provides a foundational basis for analysis, it often lacks the flexibility to integrate prior knowledge, which can constrain its effectiveness in adaptive settings. In contrast, Bayesian methods enable continual refinement of statistical inferences through the assimilation of accumulating evidence, thereby supporting more informed decision-making and improving the reliability of trial findings. This paper also considers persistent challenges in clinical investigations, including replication difficulties and the misinterpretation of statistical results, suggesting that Bayesian strategies may offer a path toward enhanced analytical robustness. Moreover, discrete probability models, specifically the Binomial, Poisson, and Negative Binomial distributions are explored for their suitability in modeling clinical endpoints, particularly in trials involving binary responses or data with overdispersion. The discussion further incorporates Bayesian networks and Bayesian estimation techniques, with a comparative evaluation against maximum likelihood estimation to elucidate differences in inferential behavior and practical implementation.


\section{Introduction.}
Over the past hundred years, the field of medical science has experienced profound developments that have revolutionized the discovery and formulation of pharmaceutical agents. These advancements have reshaped conventional medical practices, driving home the necessity for rigorous scientific methodologies to validate the safety, therapeutic efficacy, and overall reliability of medical interventions. A particularly pivotal moment in this evolution was marked by the thalidomide crisis in the late 1950s \citep{kim2011thalidomide,maity2025bayesian}. This tragedy, which led to widespread birth defects due to insufficient pre-market safety assessments, served as a wake-up call to the medical and scientific communities. It underscored the indispensable role that statistical methods must play in safeguarding public health. As a consequence, the role of statisticians in clinical research underwent a fundamental transformation. Once positioned mainly as gatekeepers who prevented erroneous interpretations of favorable results, statisticians began to take on a more integrated and collaborative role across all phases of medical investigations \citep{hertweck2023clinicopathological,khan2023myb,parkes20251519mo}.

This transition marked the beginning of an era in which statisticians became central contributors to the planning, implementation, and interpretation of clinical studies. Over the last five decades, their involvement has been critical in refining the design of clinical trials and improving the assessment of biomedical hypotheses \citep{lesaffre2020bayesian,maity2025cumulative,guo2025optimization}. Much of this transformation can be attributed to the foundational work of legendary statisticians like R.A. Fisher, Bradford Hill, Jerzy Neyman, Egon Pearson, and William Cochran. Their conceptual and methodological innovations established a robust framework that continues to underpin contemporary statistical applications in medicine. These contributions not only enhanced the scientific rigor of clinical trials but also facilitated a deeper understanding of biological mechanisms, thereby accelerating the development of life-saving medical treatments \citep{harrington2025open}. Today, the legacy of these pioneers is reflected in the widespread adoption of methodologies such as randomized controlled trials, systematic reviews, meta-analyses, and sophisticated statistical modeling. These tools have become essential in guiding clinical decision-making and shaping evidence-based healthcare practices.

At the heart of medical knowledge development lies a carefully structured process of experimentation, which is specifically designed to evaluate hypotheses concerning the behavior of medical interventions and their therapeutic value. Each experimental study, regardless of whether its findings are deemed conclusive or preliminary, plays an important role in the iterative refinement of medical theories and practices \citep{kakkat2023cardiovascular,khan2023myb}. This disciplined and incremental approach forms the foundation of modern pharmaceutical research, ensuring that drugs are rigorously vetted before they become publicly accessible. The standard pathway for drug development begins with pre-clinical investigations, typically conducted in laboratories and animal models, aimed at identifying the safety profile and pharmacokinetics of new compounds \citep{lesaffre2020bayesian,maity2024pharmacokinetic}. Should these early studies demonstrate acceptable risk levels, the compound advances to Phase I clinical trials involving a limited number of healthy participants, primarily to assess safety, tolerability, and optimal dosing.

Following successful completion of Phase I, the investigational drug proceeds to Phase II trials, which involve a larger group of patients to explore preliminary therapeutic effects and further evaluate safety. If the outcomes continue to support efficacy and tolerability, Phase III trials are conducted. These are extensive studies involving diverse populations, designed to confirm clinical benefit, detect less common adverse effects, and compare the new treatment with existing standard therapies. Once a drug successfully passes Phase III, the comprehensive data package is submitted for review by regulatory bodies such as the U.S. Food and Drug Administration (FDA) and the European Medicines Agency (EMA) \citep{maki2025new,maity2023highest}. However, regulatory approval does not signify the end of a drug’s evaluation. Post-marketing surveillance—often referred to as Phase IV remains essential. These large-scale, real-world studies continuously monitor the long term safety, effectiveness, and broader impact of the medication on public health, thus ensuring that the benefits observed during clinical trials are sustained in diverse patient populations over time.

Despite the careful and deliberate nature of this research framework, concerns have emerged regarding the reliability of medical findings. A significant issue, known as the \emph{reproducibility crisis} \citep{baker2016reproducibility,maity2023jeffreys}, has drawn attention to the fact that a considerable portion of published medical studies cannot be successfully replicated. This problem has raised doubts about the validity of certain scientific conclusions, posing challenges to the trustworthiness of medical research. Various factors have been linked to this issue, including insufficient sample sizes, selective reporting of positive outcomes, and publication bias \citep{vikramdeo2024abstract,vikramdeo2023profiling}. Additionally, poor experimental design or inadequate controls can further compromise the accuracy of results. In response to these concerns, researchers have advocated for improved scientific practices, such as stronger data collection protocols, greater transparency in reporting both positive and negative results, and increased emphasis on replication studies \citep{lesaffre2020bayesian,ghosh2023adaptive}. Regulatory organizations like the FDA and EMA have also emphasized the importance of continued post-marketing surveillance to ensure that drugs maintain their expected safety and effectiveness when used in broader populations. While the \emph{reproducibility} crisis has exposed certain flaws in the medical research process, it has also motivated meaningful reforms aimed at enhancing scientific rigor and improving the reliability of medical advancements \citep{pramanik2022lock}.

The classical statistical framework, shaped by the independent yet somewhat opposing contributions of Ronald Fisher on one side and Jerzy Neyman and Egon Pearson on the other, has played a crucial role in bringing rigor to empirical medical research \citep{dasgupta2023frequent,hertweck2023clinicopathological,khan2024mp60}. This traditional approach introduced powerful tools such as the P-value, which became widely adopted for assessing statistical significance. However, despite its utility, the P-value is frequently misunderstood, overemphasized, and misapplied in medical research. Misinterpretations of P-values often lead researchers to draw incorrect conclusions about the strength of evidence in their studies, sometimes resulting in overstated findings. Moreover, one key limitation of classical statistical methods is their inability to explicitly integrate prior knowledge into current analyses \citep{lesaffre2020bayesian}. Scientific progress is fundamentally cumulative, relying on insights gained from both past successes and failures, yet classical statistical frameworks lack the means to formally incorporate this historical information \citep{pramanik2024bayes}. Consequently, while the classical approach has been invaluable in improving research standards, it falls short in addressing the dynamic nature of scientific learning, where prior evidence should ideally inform future investigations.

The Bayesian approach offers a powerful alternative by allowing researchers to integrate prior knowledge directly into their statistical models. For many years, Bayesian methods faced resistance from the broader statistical community, with some viewing them as unconventional or unreliable \citep{pramanik2020optimization,pramanik2023semicooperation}. This skepticism persisted despite Bayesian methods offering a flexible framework that could blend historical insights with new data, thereby strengthening the conclusions drawn from research. One major reason for this resistance was the computational burden associated with Bayesian calculations, which made practical implementation difficult for most real-world problems. As a result, Bayesian analysis remained largely theoretical and was treated as a mathematical curiosity until the early 1990s \citep{pramanik2021optimala,pramanik2021scoring}. The landscape shifted dramatically with the introduction of Markov chain Monte Carlo (MCMC) sampling techniques, which provided an efficient means to perform complex Bayesian computations. This breakthrough allowed researchers to apply Bayesian methods to practical problems across various scientific fields. Since then, Bayesian analysis has surged in popularity, gaining strong support among statisticians and earning increasing recognition in the clinical research community. Its ability to combine past evidence with new data has proven particularly useful in medical research, where prior knowledge about treatment outcomes, patient characteristics, and disease patterns can significantly enhance study outcomes and improve decision-making processes \citep{lesaffre2020bayesian}.

\section{Preliminaries.}

This section focuses on key discrete probability distributions that are particularly useful in Bayesian analysis \citep{pramanik2020motivation}. To illustrate their application, consider a vaccination program initiated to combat a new strain of the hepatitis virus within an organization. On the first day of the program, ten individuals known to have resistance to the hepatitis virus are vaccinated, and their antiserum levels are measured to assess their immune responses \citep{pramanik2024estimation,vikramdeo2024mitochondrial}. The primary goal is to estimate the probability that an individual will respond positively to the vaccination. Determining this probability is crucial for understanding the effectiveness of the vaccine, especially in a population with known resistance. In Bayesian analysis, selecting the appropriate probability model is essential for making accurate inferences about such probabilities. The models that are most suitable for addressing this type of question fall under the category known as discrete probability distributions. These distributions are particularly useful when dealing with count data or binary outcomes, both of which are common in medical research scenarios like this one \citep{pramanik2024motivation}. By carefully selecting the right model, researchers can improve the accuracy of their estimates and draw meaningful conclusions about vaccine efficacy.

In this context, three important discrete probability distributions are particularly relevant: the Binomial distribution, the Poisson distribution, and the Negative Binomial distribution. Each of these models offers unique strengths that make them suitable for different types of data \citep{bulls2025assessing,pramanik2024bayes}. The Binomial distribution is ideal for modeling scenarios where there are a fixed number of trials, each with two possible outcomes success or failure making it well-suited for estimating the probability of a positive vaccine response among a defined group of participants. The Poisson distribution, on the other hand, is particularly effective for modeling the number of events that occur within a fixed interval of time or space, such as counting the number of vaccine responses observed over multiple trial periods. Meanwhile, the Negative Binomial distribution extends the Binomial framework by allowing for variability in the number of trials required to achieve a certain number of successes, making it useful for analyzing data with \emph{overdispersion}. Each of these distributions plays a crucial role in Bayesian analysis, offering flexible tools for modeling and estimating probabilities in medical research settings. By leveraging these discrete distributions appropriately, researchers can gain deeper insights into vaccination outcomes, improving their ability to predict individual responses and assess the overall effectiveness of immunization programs \citep{pramanik2023cont,pramanik2024estimation}.

\subsection{Uniform Distribution.}

A discrete random variable $Z$ is said to follow a discrete uniform distribution if it can assume precisely $M$ distinct values, each with an equal probability of occurring. This implies that no particular outcome is more likely than another. The probability of $Z$ taking a specific value $m$ among these $M$ possibilities is represented as $P(Z = m) = 1/M$. The discrete uniform distribution is considered one of the most fundamental probability distributions and is often used as a reference model or null hypothesis in statistical studies. Since all values occur with equal probability, the expected value, or mean, of $Z$, denoted by $\E[Z]$, is simply the average of all possible values, which is calculated as $(M+1)/2$ when the values span from 1 to $M$. Additionally, the variance of $Z$ is given by $(M^2 - 1)/12$, reflecting the even spread of outcomes while accounting for some degree of variability. The uniform distribution has extensive applications in multiple disciplines. In computer science, it is essential for random number generation algorithms, while in cryptography, it aids in secure key creation. It is also widely used in game theory and decision-making frameworks that require fairness \citep{pramanik2024estimation1,yusuf2025prognostic}. A classic real-world example is rolling a fair six-sided die, where each face has an equal chance of appearing, corresponding to $M = 6$. This distribution also plays a crucial role in statistical sampling, ensuring unbiased selection when choosing individuals randomly from a population \citep{pramanik2023cmbp,yusuf2025predictive}. Though achieving perfect uniform randomness in practice may be challenging due to potential biases in random number generators or physical limitations, the theoretical foundation of the uniform distribution remains a vital part of probability theory. It effectively models situations where all outcomes have equal likelihood and serves as a building block for more advanced statistical models and inferential methods.

\begin{example}
Consider a clinical trial where researchers need to randomly assign 100 patients into five treatment groups. Each patient has an equal probability of being placed in any of the five groups. The assignment follows a discrete uniform distribution since each group is equally probable.

Let $Z$ be the group assignment, where $Z \in \{1,2,3,4,5\}$. The probability of a patient being assigned to any specific group is
\begin{equation*}
\P(Z = m) = \frac{1}{5}, \quad m = 1,2,3,4,5.
\end{equation*}
To determine the expected number of patients in each group, we use 
$\E[Z] = 100/5 = 20$. Therefore, on average, each group should contain 20 patients. The variance is 
\begin{equation*}
\text{Var}(Z) = \frac{1}{12}\left(5^2 - 1\right) = \frac{24}{12} = 2.
\end{equation*}
This indicates minimal variation in group sizes under ideal conditions. Thus, the discrete uniform distribution provides a robust framework for ensuring equal probability in randomized assignments, making it an essential tool in biostatistics. This case study highlights its importance in clinical trial designs, ensuring fairness and minimizing bias.
\end{example}

\subsection{Binomial Distribution.}

When a trial yields one of two possible results, typically referred to as success and failure, the probability distribution that describes the number of successes over multiple trials is known as the binomial distribution. Consider a scenario where $n$ independent trials are conducted, with each trial having a probability $p$ of success and a probability $(1 - p)$ of failure. Let $Z$ be the total count of successful outcomes across these $n$ trials. The probability of obtaining exactly $x$ successes is determined by the binomial probability mass function (pmf), mathematically 

\begin{equation}\label{equation_binomial_pmf}
P(Z = m) = {n \choose m} p^m (1 - p)^{n - m}, \quad m = 0, 1, \dots, n.
\end{equation}

Above equation defines the likelihood of observing precisely $m$ successes, where ${n \choose m}$ represents the binomial coefficient, calculated as $\frac{n!}{m!(n-m)!}$, which enumerates the number of different ways $m$ successes can be arranged within $n$ trials. The term $p^m$ signifies the probability of achieving $m$ successes, while $(1 - p)^{n-m}$ accounts for the probability of experiencing $n-m$ failures \citep{pramanik2021consensus,pramanik2023path}. The binomial distribution is a crucial statistical tool widely applied in various domains, including medical research, quality assurance, genetics, and reliability assessments. For instance, in clinical studies, if a treatment has an effectiveness probability of $p$ and is given to $n$ patients, the binomial distribution can be used to estimate the likelihood of achieving a particular number of successful treatments. The expected value of the binomial distribution, denoted as $\E[Z] = np$, represents the average number of successes expected over repeated trials. The variance, expressed as $\text{Var}[Z] = np(1 - p)$, measures how much the number of successes deviates from the mean \citep{forbes2011statistical}. As the trial count $n$ increases while keeping $p$ constant, the distribution progressively approximates a normal distribution, a consequence of the Central Limit Theorem. The binomial distribution also serves as the foundation for the binomial test, a statistical method used to evaluate whether the observed number of successes significantly deviates from the expected outcome under a null hypothesis \citep{pramanik2021,pramanik2022stochastic}.

\begin{definition}
	A random variable $ Z $ is said to follow a Binomial distribution with parameters $n \in \mathbb{N} $ and $ p \in [0, 1] $ if and only if the pmf of $ Z $ is given by equation (\ref{equation_binomial_pmf}) and is denoted by $ Z \overset{i.i.d}{\sim} Bin(n, p) $, where i.i.d refers to independent and identically distributed.
\end{definition}

\begin{example}\label{e1}
	Consider a cancer research institute where, on average, 10 patients visit daily—3 of whom have colon cancer and the remaining 7 have lung cancer. Now, suppose a random selection of 20 cancer patients is made, with each selection being independent and allowing for replacement. The goal is to calculate the probability that exactly 5 of the selected patients are diagnosed with colon cancer.
	
	Let $ Z $  be the total number of colon cancer patients of 20 cancer patients. Hence, $ Z \overset{i.i.d}{\sim}  Bin(20, 0.3) $. Therefore, Equation (\ref{equation_binomial_pmf}) yields $ P(Z = 5)= 0.1789$. 

\end{example}

\subsection{Poisson Distribution.}

In clinical trials, particularly during patient enrollment, it is often essential to model the number of patients present at any specific point in time. This can be effectively captured using the Poisson distribution, a discrete probability distribution commonly used to represent the number of occurrences of an event within a fixed time interval or spatial region. Let \( Z \) be the number of patients at a specific time during the enrollment period \citep{pramanik2023optimization001}. In such a case, the probability of observing exactly \( m \) patients at that moment is given by the Poisson pmf. The mathematical expression of this distribution is

\begin{equation}\label{equation_poisson_pmf}
P(Z = m) = \exp(-\lambda) \frac{\lambda^m}{m!}, \quad m = 0, 1, 2, \ldots.
\end{equation}

Here, \( \lambda \) represents the expected average number of patients at a given time, while \( m \) is the number of patients we are interested in observing. The parameter \( \lambda \) is known as the rate parameter, reflecting the anticipated count of patients based on prior observations or trial predictions. The factor \( \exp(-\lambda) \) ensures that the total probability across all possible outcomes sums to 1, making the distribution valid. The term \( m! \) in the denominator accounts for the different ways \( m \) patients can be arranged within the given time frame \citep{pramanik2025construction,pramanik2025optimal}. The Poisson distribution is particularly useful in scenarios where events such as patient arrivals occur independently and at a constant average rate.

The Poisson distribution is widely applicable in clinical trials and healthcare research, especially when the events being measured (like patient arrivals) happen randomly and independently over time or in a given space. For example, during a clinical trial, the distribution can help forecast the number of patients expected at particular times, which aids in proper scheduling and resource management. With this model, trial organizers can predict patient flow and make informed decisions regarding staff allocation, the availability of medical equipment, and the prevention of system overloads. The distribution is also used in more complex modeling tasks, such as predicting rare events like adverse reactions in trials, hospital admissions, or the incidence of uncommon diseases. In essence, the Poisson distribution allows for effective planning in clinical environments by providing an accurate prediction of patient numbers and event occurrences \citep{pramanik2024stochastic,pramanik2025stubbornness,pramanik2025dissecting}. However, it is important to note that the Poisson model assumes that the rate of occurrence is constant over time, and the events are independent, which makes it particularly suited for situations where these conditions are met.

\begin{definition}
	A random variable $ Z $ is said to follow a Poisson distribution with parameter $ \lambda > 0 $ if and only if the pmf of $ Z $ is given by equation (\ref{equation_poisson_pmf}) and is denoted by $ Z \overset{i.i.d}{\sim} $ P$(\lambda) $.
\end{definition}

The Poisson distribution is a fundamental probability distribution often used to model the occurrence of rare events over a fixed interval of time or space. One of its key properties is that its mean, denoted as \( \E[Z] \), and its variance, \( V[Z] \), are identical and equal to the parameter \( \lambda \). This characteristic makes the Poisson distribution particularly useful for modeling situations where the average rate of occurrence is known, but the individual occurrences are random and independent. For an in-depth explanation and derivation of this property \citep{forbes2011statistical,pramanik2025factors}. Another important aspect of the Poisson distribution is its role as an approximation to the Binomial distribution under specific conditions. When the number of trials, \( n \), in a Binomial distribution is significantly large while the probability of success, \( p \), remains small, the Binomial distribution can be closely approximated by a Poisson distribution with the parameter \( \lambda = np \) \citep{pramanik2025optimal}. This approximation is particularly useful in practical applications where computing exact Binomial probabilities may be cumbersome due to large values of \( n \). The Poisson approximation is widely applied in various fields, including biology, telecommunications, and risk analysis, where the occurrence of rare events needs to be analyzed efficiently \citep{pramanik2021thesis,pramanik2016}.

\begin{exm}\label{e2}
	consider a hospital where the average number of adverse reactions to a new drug among patients in the trial has been estimated to be 2.1 per day. What is the probability that exactly 4 adverse reactions will occur on a given day?
	
	In a clinical trial setting, consider a hospital where the average number of adverse reactions to a new drug among patients in the trial has been estimated to be 2.1 per day. We want to determine the probability that exactly 4 adverse reactions will occur on a given day. Equation \eqref{equation_poisson_pmf} implies $ P(Z = 4)= 0.0992$.
	\end{exm}

\subsection{Negative Binomial Distribution.}

The Negative Binomial distribution is frequently employed in clinical trial analyses, particularly in scenarios where a treatment or intervention is administered until a predetermined number of successful outcomes is observed \citep{hua2019assessing,polansky2021motif}. Unlike the Binomial distribution, which focuses on a fixed number of trials and measures the number of successes, the Negative Binomial distribution instead tracks the number of unsuccessful attempts before reaching a specified success count. This characteristic makes it highly applicable in medical research, where treatments may need to be administered multiple times before achieving the desired therapeutic effect. Mathematically, if $Z$ represents the number of unsuccessful treatment attempts before obtaining \( \kappa \) successful responses, with the probability of success per trial being \( p \), then the pmf is given by  

\begin{equation}\label{equation_negbinom_pmf}
P(Z = m) = {{m - \kappa + 1} \choose m} p^\kappa (1 - p)^m, \quad m = 0, 1, 2, \ldots
\end{equation}

Equation \eqref{equation_negbinom_pmf} quantifies the probability of experiencing exactly \( m \) unsuccessful attempts before reaching the \( \kappa \)th therapeutic success \citep{pramanik2024estimation,pramanik2023cont}. The combinatorial component \( {{m - \kappa + 1} \choose m} \) accounts for the different sequences in which the unsuccessful and successful outcomes can occur. Since each treatment attempt is considered independent, the probability of achieving \( \kappa \) successful outcomes is represented by \( p^\kappa \), while the likelihood of encountering \( m \) unsuccessful attempts before reaching this threshold is given by \( (1 - p)^m \). The Negative Binomial model is especially useful in clinical studies involving overdispersed count data, where the variance exceeds the mean, which is commonly observed in trials assessing patient responses to new drugs, rehabilitation treatments, or behavioral interventions \citep{pramanik2025strategies,pramanik2023optimization001}. 

In clinical studies, the Negative Binomial distribution is advantageous for modeling count based data where treatment response rates vary among patients \citep{pramanik2025strategic,pramanik2025impact}. For instance, in trials evaluating a novel cancer therapy, researchers may need to determine the number of cycles of chemotherapy required before achieving tumor remission. Similarly, in studies assessing the efficacy of a new vaccine, the distribution can help estimate the number of exposures to a pathogen before a sufficient immune response is observed \citep{pramanik2024measuring,pramanik2024dependence}. The Negative Binomial distribution extends the geometric distribution, which is a special case where \( \kappa = 1 \), meaning the study is concerned with the number of failed attempts before a single successful response. Moreover, when analyzing clinical trial data, the Negative Binomial regression model is often preferred over the Poisson model due to its ability to account for variability in patient responses. The flexibility of this distribution enables medical researchers to better quantify treatment efficacy, assess risk factors associated with adverse events, and refine study designs to improve patient outcomes. By leveraging its properties, clinical scientists can derive more accurate statistical inferences, leading to improved decision-making in evidence-based medicine \citep{pramanik2024parametric,pramanik2025dissecting}.  
 
\begin{definition}
	A random variable $Z$ is considered to be a Negative Binomial distribution with parameters $ \kappa \in \mathbb{N}$ and $p \in [0,1]$ if and only if its pmf is defined as in equation (\ref{equation_negbinom_pmf}). This distribution is represented using the notation $ Z \overset{i.i.d}{\sim} \text{NB}(\kappa, p) $.
	\end{definition}

This distribution is widely used in various fields to model scenarios where the number of failures before achieving a set number of successes is of interest, especially when the trials are independent, and the probability of success remains constant. One of the key properties of this distribution is its mean, given by \( \E[Z] = \kappa(1 - p)/p \), where \( \kappa \) represents the number of successes and \( p \) is the probability of success in each trial. Additionally, the variance of the Negative Binomial distribution is \( V[Z] = \kappa(1 - p)/p^2 \), highlighting the spread or variability in the number of failures before the target successes are achieved \citep{valdez2025association,valdez2025exploring}. These properties make it an essential tool for modeling overdispersed count data, where the variance exceeds the mean. In fact, there are many real-world situations where the Negative Binomial distribution is a more appropriate fit than the Poisson distribution. While the Poisson distribution is useful for modeling rare events with a known constant mean rate of occurrence, it assumes that the variance equals the mean, which is often not the case in practice. The Negative Binomial distribution, on the other hand, accommodates data with higher variability, making it more suitable for processes where the variance is greater than the mean \citep{powell2025genomic,pramanik2025optimal1}. For instance, in fields such as healthcare, epidemiology, and quality control, where the occurrence of events such as adverse reactions, disease outbreaks, or product defects might exhibit significant variability, the Negative Binomial distribution provides a better fit than the Poisson model. By incorporating the potential for overdispersion, it offers more accurate predictions and insights into the underlying processes that govern these events, ultimately leading to more effective decision-making and resource allocation \citep{powell2026role}.

\begin{exm}\label{e3}
In a clinical trial, a new drug treatment has a 33\% chance of showing a positive response in patients (i.e., assuming the positive response is considered a success). What is the probability that exactly 10 patients will fail to respond to the treatment before the third successful response is observed?

The scenario describes the number of failures before obtaining a certain number of successes, which is exactly what the Negative Binomial distribution models. Equation equation (\ref{equation_negbinom_pmf}) yields the probability as 0.0432.
\end{exm}

\subsection{Conditional Probability.}

In the Bayesian analysis, the conditional probability plays an important role in understanding the relationship between various events or variables. This idea allows for the adjustment of probabilities as new information becomes available, making it essential in Bayesian methods, which focus on refining knowledge in response to evidence. Conditional relationships are frequently encountered in real-life scenarios, especially in medical practice \citep{kakkat2026angiotensin}. In healthcare settings, doctors base their line of questioning on the initial presentation of symptoms. For example, if a patient exhibits signs that could be linked to a particular diagnosis, the doctor will direct further questions specifically related to that condition. As the patient provides answers, the doctor will adapt the questions, either asking for more detailed information or omitting irrelevant inquiries \citep{ellington2025playmydata}. This process reflects the dynamic nature of medical decision-making, where conditional factors influence the flow of information gathering to improve diagnosis accuracy.

A practical illustration of conditional probability in clinical practice can be seen in how doctors consider gender and other demographic information when diagnosing a patient. If a male patient experiences fainting, the doctor would not typically investigate pregnancy as a possible cause due to biological constraints. In this case, the subsequent questions of the doctor  and diagnostic approach are guided by the patient’s gender and other relevant factors. This showcases how certain events or conditions are examined based on other known factors, influencing the medical professional’s reasoning and diagnosis. By understanding and applying conditional probability, doctors can focus their attention on the most probable causes of symptoms, tailoring their questioning and testing to gather relevant data. This structured approach to adjusting probabilities based on new evidence is a key component of Bayesian thinking, which aids in making more informed decisions, especially when uncertainty is present.

In clinical trials, it is important to understand key probability terms when analyzing results and making data-driven decisions. Consider two possible outcomes: \( E_1 \) and \( E_2 \). For example, \( E_1 \) might represent a patient experiencing a particular side effect, while \( E_2 \) could represent successful treatment. The notation \( P(E_1 \cup E_2) \) refers to the probability that at least one of these outcomes occurs, either the side effect, the treatment success, or both. When both events occur together, such as a patient experiencing the side effect and also benefiting from the treatment, this is denoted by \( P(E_1 \cap E_2) \). On the other hand, the probability that event \( E_1 \) does not take place (e.g., the absence of a side effect) is written as \( P\left(E_1^C\right) \). These expressions provide a framework for accurately evaluating the outcomes observed in the study.

\begin{definition}
	The likelihood of event \( E_1 \) occurring given that event \( E_2 \) has already occurred is expressed as the conditional probability \( P(E_1|E_2) \). This is defined as
	\begin{equation} \label{equation_conditional_probability}
	P(E_1|E_2) := \frac{P(E_1 \cap E_2)}{P(E_2)},
	\end{equation}
	for all $P(E_2)>0$.
\end{definition}

\begin{exm}\label{e4}
Research has suggested that long-acting injectable antipsychotics can enhance patient adherence, lower the likelihood of relapse, and potentially lead to better management of symptoms. Additionally, although more costly, some second-generation antipsychotics may be more tolerable than older alternatives. To investigate these findings further, a hospital carried out a randomized clinical trial evaluating the impact of long-acting risperidone injections on several outcomes: psychiatric hospital admissions, schizophrenia-related symptoms, quality of life, adherence to medication, side effects, and overall health care expenditures. The primary outcome of interest was psychiatric hospitalization. Table \ref{table_hospitalization} presents hospitalization statistics after a one-year follow-up, categorized by both gender and race. Based on the data, what is the probability that a subject was hospitalized, given that they are male?

To assess the probability of hospitalization among male subjects, we define \( M \) as the event that an individual is male and \( H \) as the event that the individual was hospitalized. The probability \( P(M \cap H) \) represents the likelihood that a randomly selected participant is both male and hospitalized, while \( P(M) \) denotes the probability of selecting a male from the study population. By utilizing Equation \eqref{equation_conditional_probability} we can determine the probability of hospitalization given that the subject is male. Using the values provided in Table \ref{table_hospitalization}, this probability is calculated as 0.405. This result indicates that, based on the study data, 40.5\% of male participants were hospitalized after one year of follow-up.	
\end{exm}

\begin{table}[H]
	\centering
	\caption{Data of hospitalization within 1 year by gender and race.}
	\label{table_hospitalization}
	\begin{tabular}{llcc}
		\hline
		Gender     & Race & \multicolumn{2}{l}{Hospitalization within 1 year}  \\
		&                  & Yes (H) & No  \\
		\hline
		Male (M)   & Caucasian        & 56      & 79  \\
		& African American & 56      & 80  \\
		& Other            & 5       & 13  \\
		\hline
		Female (F) & Caucasian        & 7       & 3   \\
		& African American & 6       & 9   \\
		& Other            & 1       & 1   \\
		\hline         
	\end{tabular}
\end{table}

\begin{rmk}
	It is important to note that the overall probability of hospitalization can be determined using the following formula
	 \begin{equation} \label{equation_total_prob} 
	 P(H) = P(H|M) P(M) + P(H|F) P(F).
	  \end{equation}
	   This concept is known as the law of total probability.
\end{rmk}

\begin{exm}\label{e5}
	Consider a scenario where the probability of surviving cancer over a certain time period differs based on the stage at which the cancer is diagnosed. For individuals with early-stage cancer, the survival probability is 0.80, whereas for those with late-stage cancer, the survival probability drops to 0.20. Furthermore, the distribution of cancer diagnoses is not equal between stages. In fact, the majority of cancers, specifically 90\%, are detected in the early stages, while only 10\% are identified in the later stages. Given these probabilities and the prevalence of each cancer stage, the question arises: what is the overall probability of survival for a randomly selected individual from this population? To answer this, we need to compute the total probability of survival, considering both early and late-stage diagnoses and their respective survival rates.
	
Define three events: \( S \) represents the survival event, \( ES \) refers to the early-stage cancer diagnosis, and \( LS \) refers to the late-stage cancer diagnosis. To calculate the overall probability of survival, we can apply the law of total probability, which allows us to compute the total probability of survival by considering the different stages of cancer and their respective probabilities. Specifically, the law of total probability states that the probability of survival \( P(S) \) is the sum of the conditional probabilities of survival given early-stage cancer and late-stage cancer, weighted by the probabilities of being diagnosed with each stage. This is mathematically expressed as:
	
	\[
	P(S) = P(S|ES) P(ES) + P(S|LS) P(LS).
	\]
	
	Substituting the known values, where the survival probability for early-stage cancer \( P(S|ES) \) is 0.8, the survival probability for late-stage cancer \( P(S|LS) \) is 0.2, the probability of being diagnosed with early-stage cancer \( P(ES) \) is 0.9, and the probability of being diagnosed with late-stage cancer \( P(LS) \) is 0.1, we get:
	
	\[
	P(S) = 0.8 \times 0.9 + 0.2 \times 0.1 = 0.74.
	\]
	
	Therefore, the overall probability of survival for a randomly selected individual is 0.74, or 74\%. This result incorporates both the higher survival rate for early-stage cancer and the lower survival rate for late-stage cancer, weighted by how frequently each stage is diagnosed.
	\end{exm}

\subsection{Bayes Theorem.}

The Bayesian approach, a fundamental concept in probability theory, originates from the principles established in Bayes’ Theorem, which was formulated by Thomas Bayes, an 18th-century Presbyterian minister with a strong inclination toward mathematics. Despite his groundbreaking insights, Bayes never published his work during his lifetime. Instead, his close associate, Richard Price, compiled and submitted his findings in 1763, two years after Bayes' passing, in a document titled \textit{An Essay toward a Problem in the Doctrine of Chances}. This work, which combined Bayes’ original research with Price’s own commentary, introduced the essential ideas behind what is now recognized as Bayesian inference. The core principle of Bayes’ Theorem revolves around updating existing knowledge in light of new evidence, making it a powerful statistical tool with applications across various disciplines, including medicine, economics, artificial intelligence, and machine learning. Although initially overlooked, Bayes’ ideas eventually gained recognition in the 20th century, particularly during World War II, when they were utilized for intelligence gathering and strategic decision-making. As computational power increased, Bayesian methods became essential for handling complex problems, leading to their widespread adoption in scientific research. Unlike classical statistical approaches which rely solely on fixed data frequencies, Bayesian techniques allow for the integration of prior knowledge, making them particularly valuable for scenarios involving uncertainty. Today, these methods are employed in areas such as financial forecasting, medical diagnosis, search engine optimization, and the training of artificial intelligence systems. The evolution of Bayes' Theorem from an unpublished manuscript to a cornerstone of modern science and technology illustrates the lasting impact of mathematical innovation and the role of probability in shaping decision-making in an increasingly complex world.

The theorem originates from a fundamental property of joint probability involving two events, $E_1$ and $E_2$. This property states that their joint probability can be expressed as follows:

\[
P(E_1\cap E_2) = P(E_1 | E_2) P(E_2) = P(E_2 | E_1) P(E_1).
\]

Therefore, the conditional probability of $E_2$ given $E_1$ is

\[
P(E_2 | E_1) = \frac{P(E_1 | E_2) P(E_2)}{Pr(E_1)}.
\]

A similar formulation applies to the complements of these events, denoted as \( E_1^C \) and \( E_2^C \). In clinical testing, where event $E_1$ represents the outcome of a diagnostic test and $E_2$ indicates whether a person has a specific disease, this principle allows us to express the test’s \textit{positive predictive value}, that is, the probability that an individual has the disease given a positive test result. This can be formulated as a function of the test’s \textit{sensitivity} \( P(E_1 | E_2) \), its \textit{specificity} \( P\left(E_1^C | E_2^C\right) \), and the \textit{prevalence} of the disease, \( P(E_2) \). Here, the prevalence represents the prior probability that an individual has the disease before considering test results, while sensitivity and specificity measure the test’s ability to correctly identify diseased and non-diseased individuals, respectively. Given these parameters, the positive predictive value can be understood as the updated probability that a person has the disease after accounting for the test outcome, reflecting a Bayesian approach to probability revision in the presence of new evidence. The actual statement of the Bayes theorem can be expressed as
\begin{align}  \label{equation_Bayes}
P(E_2|E_1) =& \frac{P(E_1|E_2) P(E_2)}{P(E_1)}  
=\frac{P(E_1|E_2) P(E_2)}{P(E_1|E_2) P(E_2) + P(E_1|E_2^C) P(E_2^C)}.
\end{align}
It is important to note that the denominator of the right hand side of Equation \eqref{equation_Bayes} is the direct result of the \emph{law of total probability}. Also,
\begin{align*}
P(E_1|E_2) =&  \frac{P(E_2 \cap E_1)}{P(E_1)} 
= \frac{P(E_1|E_2) P(E_2)}{P(E_1)}, \quad \text{by the definition of conditional probability}. 
\end{align*}

During Bayes’ time, mathematicians were primarily focused on determining the likelihood of events occurring based on a known data distribution and its assumed true parameters. However, Bayes was more fascinated by the reverse problem-understanding what the observed data could reveal about an unknown, continuous parameter. His insight was to apply probability principles to a scenario where the collected data was treated as evidence (i.e., event $E_1$), and the unknown parameter was treated as the quantity to be inferred (i.e., event $E_2$). He recognized that making inferences about an unknown parameter required assigning probability values to it, reflecting the idea that while the true value may never be known with certainty, some values might be more plausible than others. This reasoning naturally led to the concept of prior probabilities, where each possible value of the parameter is assigned a likelihood based on existing beliefs or knowledge. In this perspective, the parameter itself is considered stochastic, not because it inherently varies, but because probability is being used to quantify uncertainty about its actual value. When dealing with a continuous unknown parameter, this results in a prior probability density function that represents initial beliefs before examining the data. By integrating the observed data with this prior information, Bayes formulated a method for updating these beliefs, resulting in a posterior probability density function. This posterior distribution encapsulates the revised understanding of the parameter after accounting for new evidence, forming the fundamental principle of Bayes' Theorem.

\begin{example}  \label{example_Bayes_theorem}
	Assume that the overall survival probability for cancer patients is 0.74. Additionally, patients with early-stage cancer have a survival probability of 0.8, and the detection rate for early-stage cancer is 90\%. Given that a patient survived, we are interested in determining the probability that they had early-stage cancer. Define $E_1$ as the event where a patient has cancer at its early stage, and event $E_2$ corresponds to a patient who survived the cancer. Therefore, 
	\begin{align*}
	P(E_1|E_2) =& \frac{P(E_2|E_1) P(E_1)}{P(E_2)} 
	= \frac{0.8 \times 0.9}{0.74} 
	= 0.973.
	\end{align*}
%Associated SAS code is provided in the Appendix.	
\end{example}

\begin{example}  \label{example_hiv}
Imagine a new HIV test is introduced. If a person actually has HIV, the test correctly identifies it with a probability of 0.95. Conversely, the test correctly identifies someone without HIV as negative with a probability of 0.98. The prevalence of HIV in the state is 0.1\%, meaning 0.1\% of the population is HIV positive. To determine the probability that a person truly has HIV, given that they tested positive using this test, let us define the following events, $HIV+$: a person is in fact HIV positive, $HIV-$: A person is not HIV positive, $Test+$: The HIV test results in HIV positive, and $Test-$: The HIV test results in HIV negative.

We are given that 
\[
P(\text{Test+} \mid \text{HIV+}) = 0.95,
\]
which represents the \textit{sensitivity} of the test. Moreover, 
\[
P(\text{Test-} \mid \text{HIV-}) = 0.98,
\]
which is the \textit{specificity} of the test. A good diagnostic test ideally has both high sensitivity and high specificity. Based on these values, the test appears to be fairly accurate. %However, when applying Bayes' theorem implemented in the SAS code using Equation~(\ref{equation_Bayes})
The resulting probability that a person is actually HIV positive given they received a positive test result (i.e., a false positive scenario) is only about 0.045.  This implies that, despite the test’s high sensitivity and specificity, the probability that an individual who tested positive is truly infected is only around 5\%.	

\begin{align*}
P(HIV+|Test+) =& \frac{P(Test+|HIV+) P(HIV+)}{P(Test+)}  \\
=& \frac{P(Test+|HIV+) P(HIV+)}{P(Test+|HIV+) P(HIV+) + P(Test+|HIV-) P(HIV-)}  
= 0.045.
\end{align*} 
\end{example}

\begin{figure}[H]
	\includegraphics[width=1\textwidth]{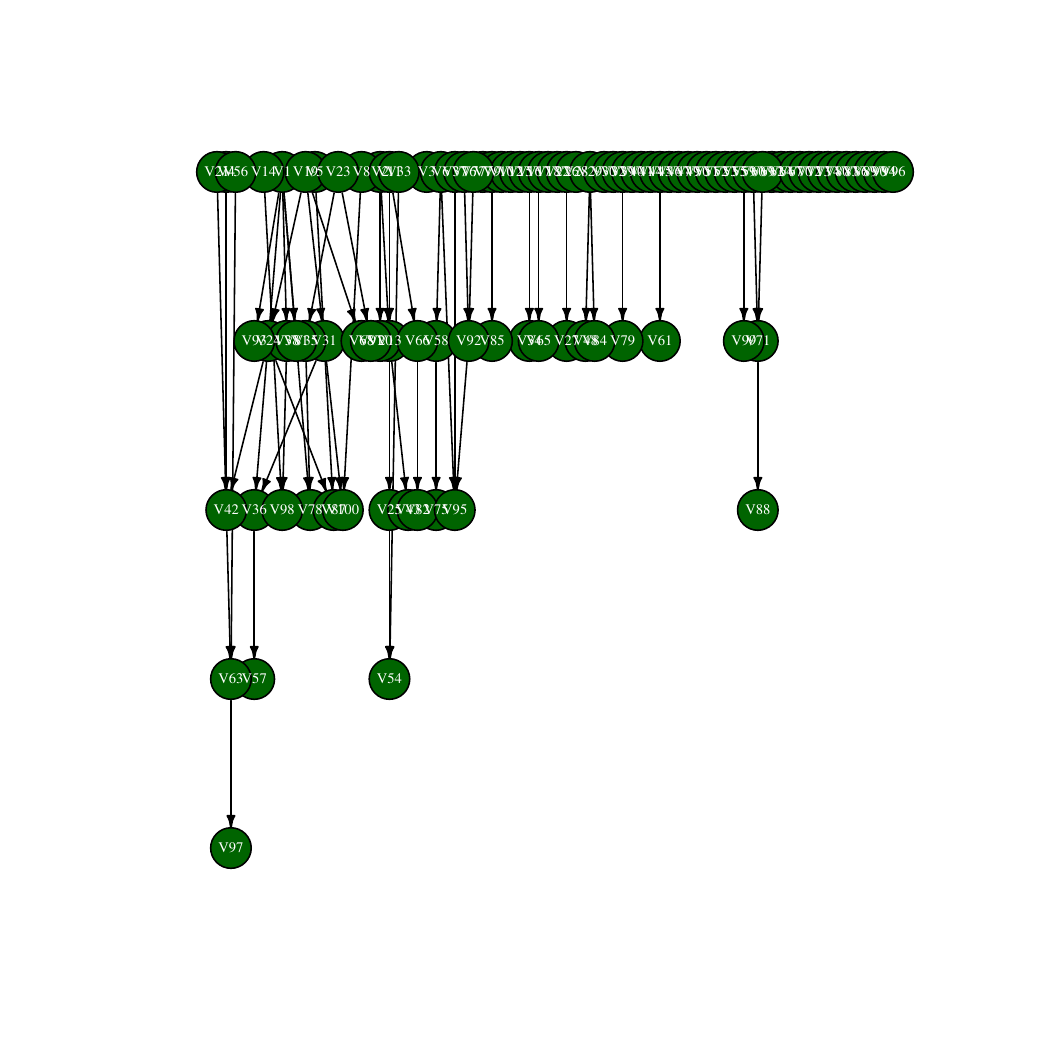}
	\caption{Bayesian Network with 100 HIV patients (Directed Acyclic Graph ).}
	\label{fig:2}
\end{figure}

\begin{figure}[H]
	\includegraphics[width=1\textwidth]{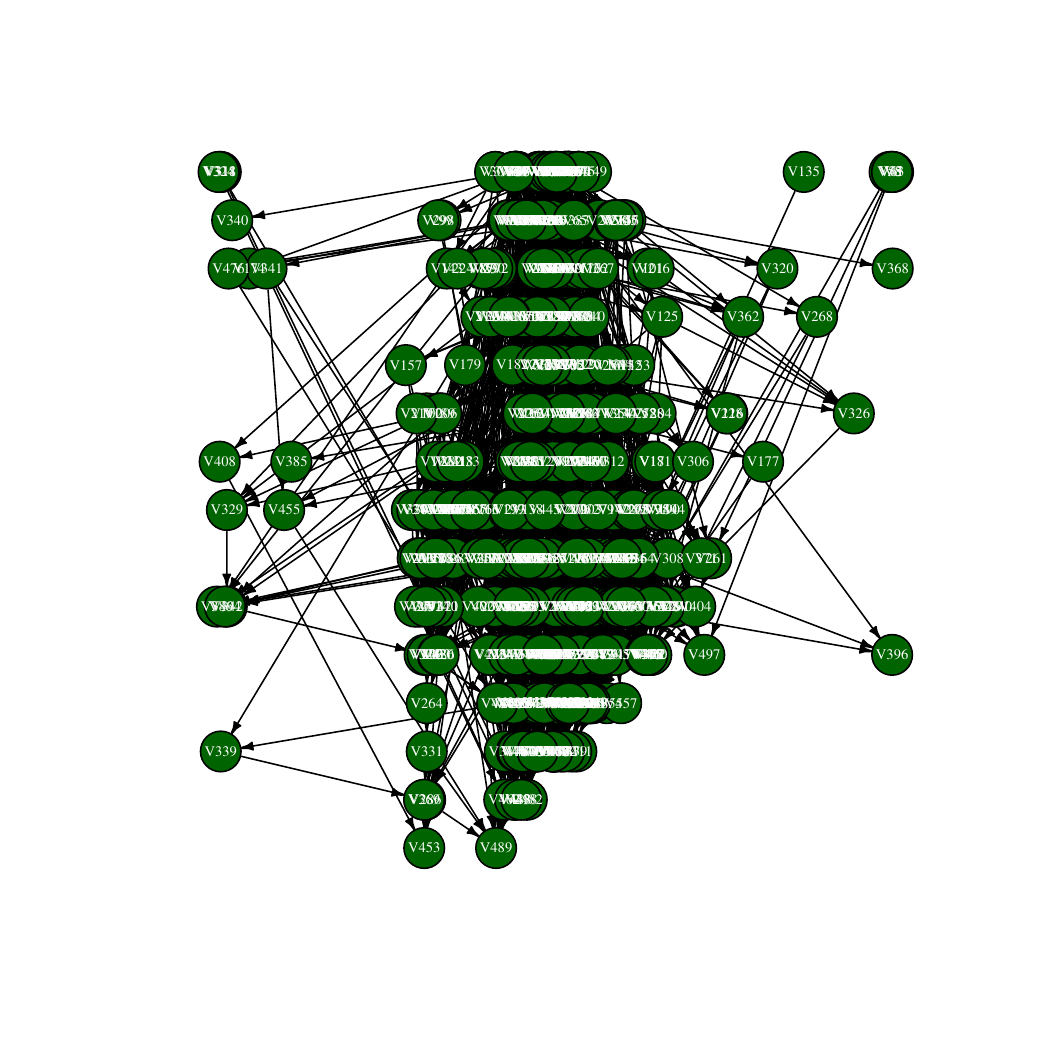}
	\caption{Bayesian Network with 500 HIV patients DAG.}
	\label{fig:3}
\end{figure}

The network generated in Figure \ref{fig:2} is a random Bayesian Network with 100 nodes, where each node represents a  patient (based on a certain characteristic), and edges represent probabilistic dependencies between these variables. The structure is a Directed Acyclic Graph (DAG), ensuring no feedback loops and preserving a flow of causal or conditional relationships. This mirrors the structure used in the original HIV testing example, where two nodes - HIV-status and Test-result  were connected by a directed edge representing how the probability of a test result depends on the underlying HIV status. The larger 100-node DAG is an expanded version of this idea, simulating a more complex real-world setting where many factors might interact probabilistically (e.g., symptoms, diagnostics, behaviors), enabling probabilistic inference across the network. Both cases exemplify the Bayesian reasoning approach using observed data (like a test result) to update beliefs about unobserved variables (like true infection status) based on the network structure. Figure \ref{fig:3} considers the same structures with 500 patients.

\begin{figure}[H]
	\includegraphics[width=1\textwidth]{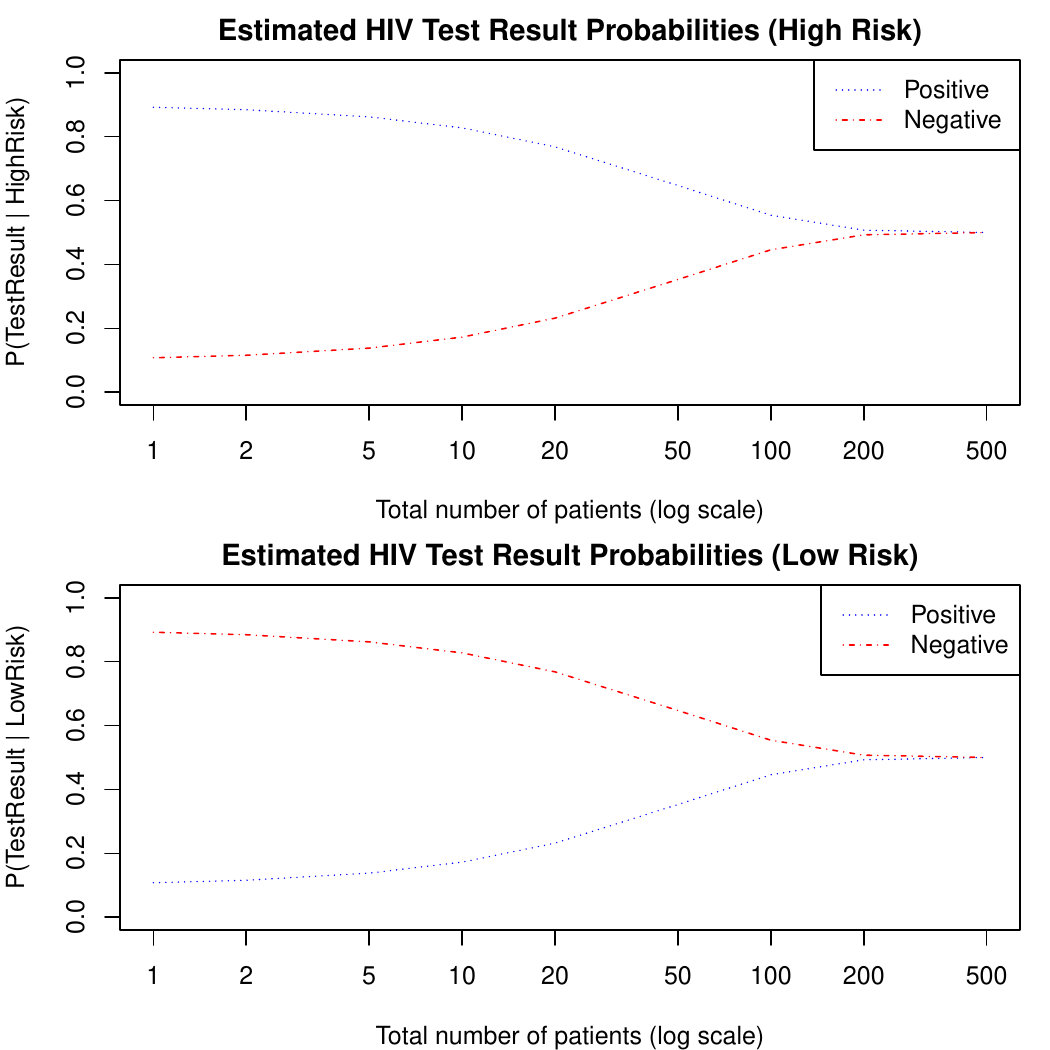}
	\caption{Estimated HIV test result probabilities converge to a uniform distribution as prior strength increases, illustrating the diminishing influence of risk classification under strong prior assumptions.}
	\label{fig:4}
\end{figure}

Figure~\ref{fig:4} presents a simulated Bayesian analysis of HIV test outcomes, illustrating how prior beliefs can influence estimated probabilities under varying strengths of prior information. In this scenario, individuals are classified into two categories based on their HIV risk profile: HighRisk and LowRisk. The goal is to examine how the estimated probabilities of testing Positive or Negative evolve as the weight of prior assumptions increases represented by the ``total number of patients" on the horizontal axis, plotted on a logarithmic scale. The two panels show results for each risk group separately.

In the top panel, individuals in the HighRisk group are initially assumed to have a high probability (90\%) of testing positive. As the imaginary sample size increases, this probability gradually decreases and converges to 50\%, reflecting a uniform distribution between Positive and Negative outcomes. This trend demonstrates how dominant priors—if not updated with real data which tend to neutralize the model’s discriminatory power, ultimately yielding uninformative estimates. Conversely, in the bottom panel, LowRisk individuals begin with a low assumed probability (10\%) of testing positive. As the prior becomes stronger, this too drifts toward a 50-50 distribution, mirroring the behavior observed in the HighRisk group. This convergence to a uniform distribution in both panels illustrates a fundamental principle in Bayesian analysis: when priors are overly emphasized relative to data (i.e., when imaginary sample sizes are large), the model's ability to distinguish between different groups diminishes. The dotted and dash-dot lines in each plot correspond to the estimated probabilities of Positive and Negative test results, respectively, making the pattern of convergence visually clear. Overall, this simulation underscores the importance of balancing prior assumptions with empirical evidence, especially in sensitive domains like HIV testing, where both overconfidence and underconfidence in model predictions can have significant real-world implications.	

\subsection{Bayes Factor.}

The concept of \emph{odds} is a useful way to express the likelihood of an event relative to its complement, offering a more intuitive interpretation in many real-world scenarios. Rather than simply stating the probability that an event will occur, odds describe how many times an event is expected to happen compared to how many times it is not. Specifically, the odds of an event \( E_2 \) are defined as the ratio of the probability that \( E_2 \) occurs to the probability that \( E_2^C \) (its complement) occurs. This can be interpreted as the ratio of favorable outcomes to unfavorable outcomes. For instance, if a person says the odds of success are ``3 to 2," this implies that for every 3 successful outcomes, there are 2 failures, yielding odds of \( 3:2 \). A practical example might involve rolling a standard six-sided die: the probability of rolling a 5 or a 6 (i.e., rolling a number 5 or greater) is \( \frac{2}{6} \), and the probability of not rolling a 5 or 6 is \( \frac{4}{6} \), leading to odds of \( \frac{2}{4} \) or \( 1:2 \) after simplification.

Building on this understanding, consider a situation in which we are interested in the odds of one event, say \( E_2 \), occurring given that another event, \( E_1 \), has already occurred. Specifically, we wish to compute the \textit{posterior odds}: 
\[
\frac{P(E_2 \mid E_1)}{P(E_2^C \mid E_1)}.
\]
This represents the updated belief in the likelihood of \( E_2 \) occurring relative to \( E_2^C \), now that we know \( E_1 \) has taken place. This formulation is especially important in fields such as medical diagnostics, machine learning, and decision theory, where we often need to update our belief in a hypothesis after observing some evidence. Bayes’ theorem provides a powerful framework for calculating this ratio, 
\begin{equation} \label{equation_Bayes_odds}
\frac{P(E_2 \mid E_1)}{P(E_2^C \mid E_1)} = \frac{P(E_1 \mid E_2)}{P(E_1 \mid E_2^C)} \times \frac{P(E_2)}{P(E_2^C)}.
\end{equation}
Equation \eqref{equation_Bayes_odds}, often referred to as the \textit{Bayes odds formula}, transforms complex conditional relationships into a product of more accessible quantities. It enables us to incorporate both prior beliefs and new empirical evidence into a coherent posterior belief.

Each component of equation \eqref{equation_Bayes_odds} carries significant meaning. On the right-hand side, the first term \( \frac{P(E_1 \mid E_2)}{P(E_1 \mid E_2^C)} \) captures the \textit{likelihood ratio}, or \textit{empirical evidence} how much more likely we are to observe \( E_1 \) if \( E_2 \) is true compared to if \( E_2^C \) is true. The second term, \( \frac{P(E_2)}{P(E_2^C)} \), is known as the \textit{prior odds}, which reflects our belief about the likelihood of \( E_2 \) before observing \( E_1 \). Finally, the left-hand side of the equation, \( \frac{P(E_2 \mid E_1)}{P(E_2^C \mid E_1)} \), represents the \textit{posterior odds}, or our updated belief about the likelihood of \( E_2 \) after taking \( E_1 \) into account. The power of this formulation lies in its ability to distill the process of belief updating into a single, interpretable ratio. It makes clear how new data (evidence) modifies our previous assumptions (priors) to yield new conclusions (posteriors), making it an essential tool in rational decision-making under uncertainty.

The term \textit{empirical evidence} in the Bayes odds formulation is more formally known as the \textit{Bayes factor}, or sometimes as the \textit{likelihood ratio}. This quantity plays a pivotal role in Bayesian inference, providing a rigorous and interpretable measure of how strongly observed data support one hypothesis over another. The Bayes factor is firmly rooted in statistical theory and has a long standing presence in the literature, supported by both theoretical justification and practical application. It serves to quantify the strength of evidence in favor of a particular scientific proposition, hypothesis, or model, offering a direct and meaningful way to evaluate how new data shift our beliefs. The concept was initially introduced by the eminent statistician Sir Harold Jeffreys (1935) \citep{jeffreys1935some}, and was further developed and extensively discussed in his book \cite{jeffreys1998theory}. Jeffreys' work laid the foundation for what is now considered one of the cornerstones of Bayesian statistical reasoning. For readers interested in a deeper understanding of the historical development, theoretical foundations, and broad applications of the Bayes factor in scientific research, the article by Kass and Raftery (1995) \citep{kass1995bayes}, offers an exceptional overview of the evolution of the Bayes factor and highlights its significance across diverse disciplines.

From a mathematical standpoint, the \textit{posterior odds} of event \( E_2 \), after accounting for observed evidence, can be expressed as the product of the \textit{prior odds} of event \( E_2 \) and the \textit{Bayes factor}. In other words, the Bayes factor is the ratio of the posterior odds of \( E_2 \) to its prior odds, serving as a multiplier that adjusts our initial beliefs in light of new information. What makes the Bayes factor particularly powerful is that it encapsulates the influence of the data independently of any specific prior belief-regardless of the magnitude of the prior odds, the Bayes factor remains an objective measure of evidence provided by the data alone. This property makes it extremely valuable for updating beliefs in a principled and consistent manner. Furthermore, in the context of hypothesis testing, the Bayes factor is frequently viewed as a Bayesian alternative to the classical \textit{frequentist} \( p \)-value. While a \( p \)-value indicates how surprising the observed data would be under a null hypothesis, it does not directly measure support for competing hypotheses. In contrast, the Bayes factor offers a clear and coherent quantification of the relative evidence for two competing models or hypotheses, making it a more informative and interpretable tool for decision-making in scientific research and statistical inference.

\begin{example}[Continuation of Example \ref{example_hiv}] \label{example_hiv0}
Assume we are interested in testing the null hypothesis \( H_0 \): that a randomly chosen individual is HIV positive, against the alternative hypothesis \( H_1 \): that the individual is not HIV positive. In this context, the result of the HIV diagnostic test serves as the observed data. Suppose the test result indicates that the individual is HIV positive. Given this outcome, we are interested to interpret the result of the hypothesis test using the Bayes factor. The Bayes factor is 
\begin{align*}
BF =& \frac{P(Test+|HIV+)}{P(Test+|HIV-)}  = 47.5.
\end{align*}
This suggests that, based on the test data indicating the individual is HIV positive, there is strong evidence supporting the validity of the null hypothesis.	
\end{example}

\begin{figure}[H]
	\includegraphics[width=1\textwidth]{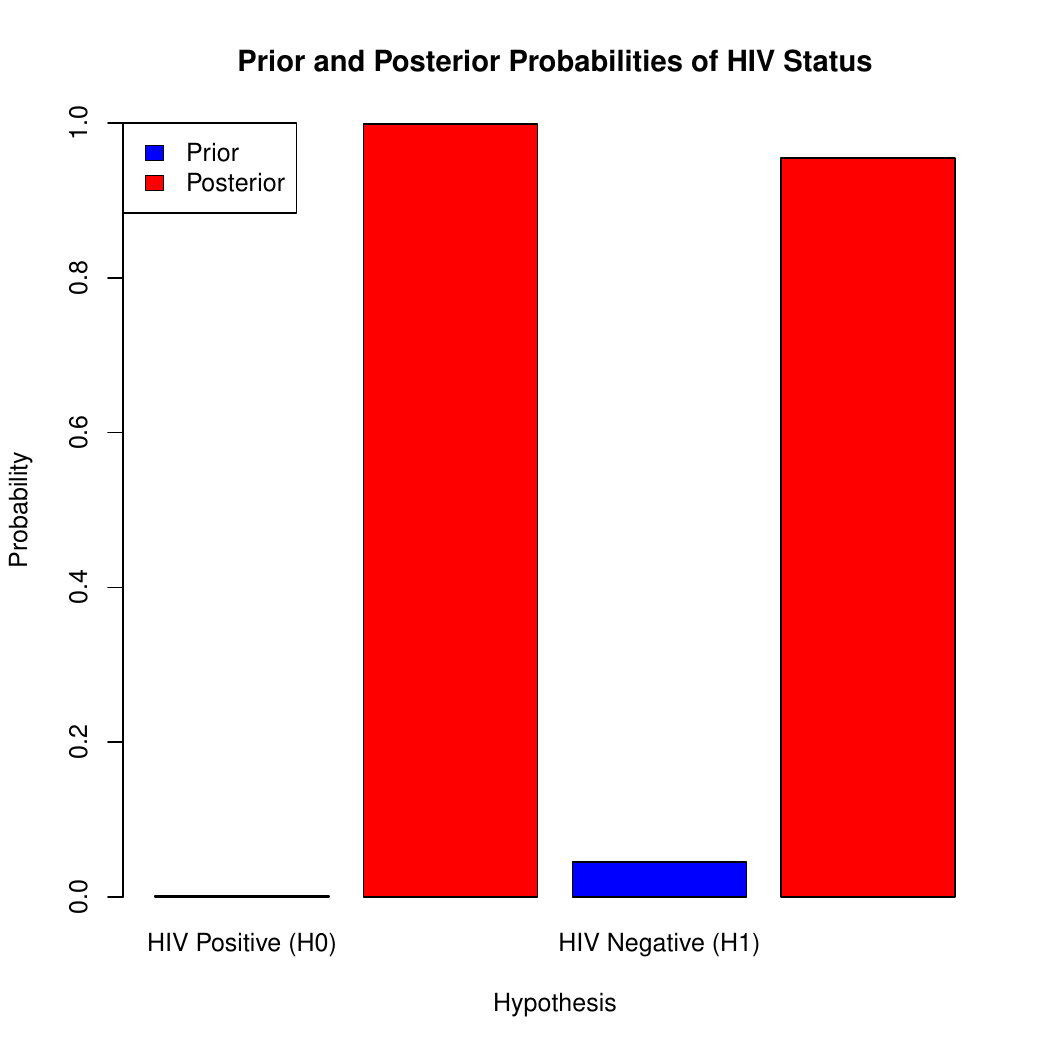}
	\caption{Updating Beliefs with Bayes Factor: Prior vs Posterior Probabilities of HIV Status.}
	\label{fig:1}
\end{figure}

Figure \ref{fig:1} presents a side-by-side comparison of the prior and posterior probabilities associated with the hypotheses that a randomly chosen individual is HIV positive ($H_0$) or HIV negative ($H_1$). The prior probabilities are based solely on the prevalence of HIV in the population, which is relatively low (0.1\%), resulting in a prior probability of 0.001 for being HIV positive and 0.999 for being HIV negative. Upon receiving a positive test result from a diagnostic tool with a Bayes factor of 47.5, these prior beliefs are updated to reflect the strength of the new evidence. The posterior probabilities, computed using Bayes' theorem, now show a notable increase in the probability of being HIV positive, despite the initial low prior. This shift highlights the substantial influence of a highly informative test result on belief updating, as the posterior probability of being HIV positive becomes considerably larger than the prior. The bar plot visually illustrates this update, reinforcing the central idea in Bayesian inference that strong empirical evidence quantified here by the Bayes factor can significantly reshape our confidence in a hypothesis even when the prior belief is weak.

However, when the prior odds of being HIV positive, calculated as \( 0.001/(1 - 0.001) \), are multiplied by the Bayes factor, the resulting posterior odds of being HIV positive given the test data is 0.0475. This outcome aligns with the conclusion presented in Example \ref{example_hiv}.

\section{Bayes Factors and $p$-values.}

The Bayes factor serves as a powerful statistical tool that quantifies the strength of evidence provided by the data in favor of one scientific theory or hypothesis over another. It is a ratio that compares the likelihood of the data under two competing models or hypotheses, with one often being $H_0$. In hypothesis testing, the Bayes factor evaluates the evidence against $H_0$. Small values of the Bayes factor suggest evidence against $H_0$, whereas large values indicate stronger evidence in favor of $H_0$. This contradicts with the frequentist approach, which utilizes a $p$-value to determine statistical significance. In this paradigm, the $p$-value represents the probability of obtaining a result at least as extreme as the observed data, assuming the null hypothesis is true. If the $p$-value falls below a predetermined threshold, usually 0.05, $H_0$ is rejected. However, the Bayes factor provides a more nuanced interpretation, allowing researchers to assess the magnitude of evidence for or against the null hypothesis rather than merely providing a binary decision of acceptance or rejection.

While both the Bayes factor and $p$-value are used in hypothesis testing, there are key conceptual differences that distinguish the two methods. One of the primary distinctions lies in how the two quantities are calculated. The $p$-value is derived by evaluating the probability of obtaining a result at least as extreme as the observed data, assuming $H_0$ is true. This means the $p$-value is calculated under the assumption of $H_0$, offering insight into the rarity of the observed outcome under that hypothesis. In contrast, the Bayes factor is computed directly from the data, comparing the relative likelihood of two competing models or hypotheses. Another notable difference is the range of values that each measure can take. As a probability, the $p$-value is constrained between 0 and 1, where smaller values suggest stronger evidence against the null hypothesis. On the other hand, the Bayes factor is a ratio of two positive quantities, meaning it can range from 0 to infinity. In a logarithmic scale, the Bayes factor can span from negative infinity to positive infinity, offering a more expansive and continuous measure of evidence. This fundamental difference in the nature of the two measures underscores the broader interpretive flexibility that the Bayes factor provides, making it a valuable tool for statistical inference and hypothesis testing.

Suppose a Bayes factor is computed as 1/10, meaning that the probability of the observed data under the null hypothesis is one-tenth the probability of the data under the alternative hypothesis. This numerical result has several equivalent interpretations, all of which convey the same fundamental shift in evidential weight. First, it implies that the data are 10 times more probable under the alternative hypothesis than under the null hypothesis, providing a strong indication in favor of the alternative. Second, it means that the observed data support the alternative hypothesis 10 times as strongly as they support the null hypothesis, emphasizing the relative strength of the evidence in favor of the alternative. Finally, in terms of Bayesian updating, it indicates that the posterior odds of the null hypothesis relative to the alternative are now one-tenth of what they were before observing the data; in other words, the Bayes factor has decreased the prior odds in favor of the null by a factor of 10. Taken together, these interpretations illustrate how a Bayes factor of 1/10 quantifies a substantial shift in belief toward the alternative hypothesis in light of the observed data.

For those who are accustomed to the threshold-based \( p \)-value approach commonly used in classical hypothesis testing, the decision-making process typically involves comparing the computed \( p \)-value to a predefined significance level, often 0.05. In this framework, if the \( p \)-value falls below the 0.05 threshold, the null hypothesis is rejected in favor of the alternative, indicating that the observed data would be highly unlikely under the assumption that the null hypothesis is true. A similar threshold-based reasoning can be applied when using Bayes factors in a Bayesian framework, though the interpretation is conceptually different. Rather than focusing on the improbability of the data under the null hypothesis alone, the Bayes factor evaluates the relative likelihood of the observed data under both the null and alternative hypotheses. A large Bayes factor provides evidence in favor of the null hypothesis, suggesting that the data are much more likely under the null than under the alternative. Conversely, a small Bayes factor supports the alternative hypothesis. This shift from a binary reject-or-fail-to-reject decision to a graded measure of evidence allows for a more nuanced interpretation of results. The commonly accepted thresholds that guide these interpretations, such as what constitutes weak, substantial, strong, or decisive evidence are summarized in Table~\ref{table_Bayes_factor}, offering a practical framework for researchers to make evidence-based decisions using Bayes factors in a manner analogous to, yet more informative than, the traditional \( p \)-value approach.

\begin{table}[H]
	\centering
	\caption{Bayes factor thresholds to find the evidence.}
	\label{table_Bayes_factor}
	\begin{tabular}{|ll|}
		\hline
		Bayes factor & Evidence for the null hypothesis  \\
		\hline
		1 to 3.2     & Not worth more than a bare mention  \\
		3.2 to 10    & substantial                         \\
		10 to 100    & Strong                              \\
		$>$ 100      & Decisive                            \\
		\hline
	\end{tabular} 
\end{table}

It is important to consider twice the natural logarithm of the Bayes factor when interpreting the strength of evidence in favor of one hypothesis over another, as this transformation places the Bayes factor on a scale that is directly comparable to familiar statistical measures such as the deviance and the likelihood ratio test statistics. These classical statistics are often expressed on a log-likelihood scale, so converting the Bayes factor in this way facilitates an intuitive and coherent interpretation within the broader context of statistical inference. By rounding and adopting a stringent thresholdm specifically, using 20 rather than 10 as the benchmark for what constitutes strong evidence, we arrive at a modified classification scheme that aligns closely with conventional standards in hypothesis testing. This adjustment enhances the practical utility of the Bayes factor, allowing researchers to assess evidence strength using a familiar numerical framework and interpretive thresholds. The resulting categorization, presented in Table \ref{table_log_Bayes_factor}, offers a refined and accessible reference for evaluating the weight of evidence in Bayesian analyses, especially in cases where alignment with frequentist statistics improves communication or understanding.

\begin{table}[H]
	\centering
	\caption{Bayes factor thresholds to find the evidence.}
	\label{table_log_Bayes_factor}
	\begin{tabular}{|lll|}
		\hline
		$ 2 \log_e$(Bayes factor) & Bayes factor & Evidence for the null hypothesis  \\
		\hline
		0 to 2  & 1 to 3     & Not worth more than a bare mention  \\
		2 to 6  & 3 to 20    & substantial                         \\
		6 to 10 & 20 to 150  & Strong                              \\
		$>$ 10  & $>$ 150    & Decisive                            \\
		\hline
	\end{tabular} 
\end{table} 
Empirically, Bayes factors have been shown to provide appropriate and practical guidelines for evaluating evidence in statistical inference, as discussed in detail by \cite{kass1995bayes}. The magnitude of a Bayes factor plays a crucial role in determining the strength of evidence in favor of one hypothesis over another. However, it is important to recognize that even an extreme Bayes factor may not be sufficient to overturn a strong prior belief. This interaction between prior belief and new evidence is a central feature of Bayesian analysis and highlights its departure from purely data-driven inference. For example, in Example~\ref{example_hiv}, the Bayes factor for a positive HIV test result is calculated to be 47.5, which by standard interpretation represents strong evidence that the individual is HIV positive. Nevertheless, when this Bayes factor is combined with the prior belief which in this case reflects the very low base rate of HIV in the general population, the resulting posterior probability of being HIV positive, even after a positive test, is only about 0.0475. This example underscores the powerful influence of prior information in Bayesian reasoning: a very low prior probability can effectively dampen the impact of even strong evidence from the data. It serves as a critical reminder that the interpretation of evidence in a Bayesian context must always take into account both the strength of the new data and the plausibility of the hypotheses before the data were observed.

In essence, while Bayesian theory is fundamentally rooted in the process of updating prior beliefs with observed data to arrive at a posterior distribution, the influence of the prior can sometimes overshadow the contribution of the data, particularly when the prior is strong or highly informative. Ideally, the Bayesian framework allows for a balanced synthesis of prior knowledge and empirical evidence, enabling parameter estimation to reflect both sources of information. However, when a prior distribution is overly dominant, it can prevent meaningful insights from the data from fully influencing the posterior results, potentially leading to biased or misleading conclusions. This dual dependence on both data and prior belief is a distinctive feature of Bayesian inference, setting it apart from purely frequentist approaches. While some statisticians, such as Bradley Efron, have criticized this feature as a fundamental weakness, arguing that it introduces subjectivity and reduces the objectivity of statistical conclusions; others, like Dennis Lindley, have embraced it as one of the strengths of being a statistician, allowing for the integration of substantive domain knowledge into the analysis. The ongoing debate has prompted the development of objective Bayesian methods, which aim to minimize the influence of the prior by employing non-informative or reference priors, thereby preserving the integrity of data-driven inference. Notably, James O. Berger and his colleagues have contributed extensively to this area, producing a rich body of literature that explores the theory and application of objective Bayesian analysis. For those interested in a broader historical perspective on the evolution of Bayesian thought and the philosophical debates surrounding it, an excellent retrospective can be found in \citep{lehmann2008reminiscences}, which offers valuable insights into how these ideas have shaped modern statistical thinking.

\section{Bayes Theorem in Discrete Distributions.}

\begin{figure}[H]
	\centering
	\includegraphics[height = 2 cm, width = 7 cm]{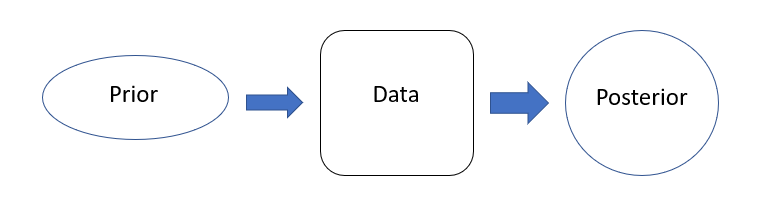}
	\caption{The Bayes rule.}
	\label{figure_bayes_rule}
\end{figure}

When conducting a Bayesian analysis, one of the key outcomes is the estimation of model parameters through what is known as the posterior distribution. The term posterior arises because it represents our updated belief about the parameters after combining prior information with the observed data. This process of updating is mathematically grounded in Bayes' theorem, which serves as the foundation of the Bayesian framework. Figure~\ref{figure_bayes_rule} provides a schematic overview of how this methodology is employed in practice. The process begins by specifying a statistical model for the observed data, typically through a likelihood function. The likelihood is derived from the probability distribution of a random variable that captures how the data are expected to behave under different parameter values. These parameters, in contrast to the fixed nature assumed in frequentist statistics, are treated as random variables in the Bayesian paradigm, which allows for the incorporation of prior beliefs about their possible values.

These prior beliefs are encoded through a prior distribution, which expresses the analyst's knowledge or assumptions about the parameters before seeing the data. The combination of this prior distribution with the data-driven likelihood function yields the posterior distribution via Bayes’ theorem. This posterior distribution is a complete summary of what is known about the parameters after accounting for both the prior and the data. From the posterior, one can extract useful summaries such as the posterior mean, median, mode, credible intervals, and other measures to draw inferences and make decisions. In many practical applications, the posterior mean is used as a point estimate of the parameter, as it reflects an average belief about its value after considering all available information. The Bayesian framework, through this cycle of prior, likelihood, and posterior, not only allows for a coherent updating of knowledge but also offers a flexible and principled way to incorporate expert opinion, historical data, or subjective uncertainty into statistical modeling.

It can be shown that the frequentist penalized likelihood maximization problem is equivalent to obtain the posterior modes when using the penalty term as the prior distribution.

\begin{example}  \label{example_coin_toss}
	Two different HIV tests are administered 10 times each on individuals known to be HIV positive. For Test A, 6 out of 10 tests return positive results. For Test B, 4 out of 10 tests return positive results. It is known a priori that the sensitivity (i.e., the probability that the test correctly returns a positive result for an HIV-positive individual) for each test is either 0.35 or 0.5.
	
	To derive the posterior we begin by denoting \( \theta \) as the \textbf{sensitivity} of the HIV test, i.e., the probability that the test returns a positive result for a person who is truly HIV positive. Let \( Z_1 \) and \( Z_2 \) be the number of positive results obtained from administering \textbf{Test A} and \textbf{Test B} respectively on known HIV-positive individuals. Then we assume
	\[
	Z_1 \sim \text{Bin}(10, \theta), \quad Z_2 \sim \text{Bin}(10, \theta).
	\]
 Test A was positive in 6 out of 10 cases, and Test B was positive in 4 out of 10 cases, under the same assumption about sensitivity \( \theta \). We denote the observed data as \( \bm{Z} = (Z_1 = 6, Z_2 = 4) \).
	
	The prior distribution on the test sensitivity \( \theta \) is:
	\[
	P(\theta = 0.5) = P(\theta = 0.35) = 0.5.
	\]
	
By Bayes' theorem, the \textbf{posterior distribution} of \( \theta \) given the observed data yields
	\[
	P(\theta|\bm{Z} = \bm{m}) = \frac{P(\bm{Z} = \bm{m}|\theta) P(\theta)}{P(\bm{Z} = \bm{m})}.
	\]
	
Using the law of total probability, and assuming that the results from Test A and Test B are conditionally independent given \( \theta \), we have
	\begin{align*}
	P(\bm{Z} = \bm{m}) &= P(Z_1 = 6, Z_2 = 4|\theta = 0.35) \cdot P(\theta = 0.35) +  P(Z_1 = 6, Z_2 = 4|\theta = 0.5) \cdot P(\theta = 0.5) \\
	&= P(Z_1 = 6|\theta = 0.35) \cdot P(Z_2 = 4|\theta = 0.35) \cdot 0.5  + P(Z_1 = 6|\theta = 0.5) \cdot P(Z_2 = 4|\theta = 0.5) \cdot 0.5 \\
	&= 0.5 \cdot \binom{10}{6}(0.35)^6(0.65)^4 \cdot \binom{10}{4}(0.35)^4(0.65)^6  +0.5 \cdot \binom{10}{6}(0.5)^{10} \cdot \binom{10}{4}(0.5)^{10}=0.033.
	\end{align*}
	
	This yields the marginal likelihood \( P(\bm{Z} = \bm{m}) \). Bayes' rule implies
	\[
	P(\theta = 0.35 | \bm{Z}) = \frac{P(Z_1 = 6|\theta = 0.35) \cdot P(Z_2 = 4|\theta = 0.35) \cdot 0.5}{P(\bm{Z} = \bm{m})}=0.151.
	\]
	\[
	P(\theta = 0.5 | \bm{Z}) = \frac{P(Z_1 = 6|\theta = 0.5) \cdot P(Z_2 = 4|\theta = 0.5) \cdot 0.5}{P(\bm{Z} = \bm{m})}=0.849.
	\]
	
	Thus, the posterior distribution reflects the updated belief about the test sensitivity \( \theta \) given the evidence from both Test A and Test B. This allows us to quantify our uncertainty about the sensitivity after observing actual test results, grounded in both prior assumptions and empirical data. %The SAS code for the marginal and posterior distributions are given in the Appendix. 
    One can compute the Bayes factor of the null hypothesis that the probability is 0.5, using the likelihood from the above output. Since Bayes factor is the ratio of likelihoods, then again using SAS yields the Bayes factor to be 5.64. This implies that there is substantial evidence in the favor of the probability to be 0.5.
	\end{example}

\begin{example}  \label{example_blood}
	Following a car accident on a nearby highway, four patients are admitted to the hospital; one with blood group A and three with blood group B. Each patient requires a blood transfusion, and the public is asked to donate blood. The probability that a random donor’s blood matches group A is 0.2, and for group B, it is 0.1. A total of seven individuals donate, and exactly one match is found. 
	
To determine the posterior probability that the successful blood type match occurred for a patient with blood group A or B, given the data from blood donations. Let us denote \( \theta \) as the probability of a match in blood types during a donation. Let \( Z \) be the number of individuals (or trials) required until the first successful match occurs. Under this setup, \( Z \) follows a Negative Binomial distribution with parameters \( (1, \theta) \), as we are modeling the number of trials needed to obtain the first success. The values of \( \theta \) differ depending on the blood group: for group A, the probability of a match is \( \theta = 0.2 \), and for group B, it is \( \theta = 0.1 \). Since we have one patient with blood type A and three with blood type B, and no explicit prior information is provided beyond these numbers, we adopt the prior probabilities \( P(\theta = 0.2) = 0.25 \) and \(P(\theta = 0.1) = 0.75 \), which reflect the relative frequencies of the blood groups needing donation. These priors account for the fact that only one out of four patients has blood group A, while the remaining three have group B. Given that a match was observed on the first donation (i.e., \( Z = 1 \)), we apply Bayes' theorem to compute the posterior probabilities. The computation, 
%which can be performed using software such as SAS (code is provided in the Appendix), 
yields the following results
	\begin{align*}
	P(\theta = 0.2 | Z = 1)& = 0.226,\\ P(\theta = 0.1 | Z = 1)& = 0.774.
	\end{align*}
	
	These posterior probabilities indicate that, even though a match occurred immediately, it is still more likely that the match was for a patient with blood group B, primarily because there were more patients with this group and thus a higher prior probability was assigned to \( \theta = 0.1 \). This example nicely illustrates how Bayesian inference combines prior information with observed data to refine our beliefs about an uncertain event, in this case, the identity of the matched blood group.
	
	Nonetheless, if one assumes a uniform discrete prior on $ \theta $ that is $ P(\theta = 0.2) = P(\theta = 0.1) = 0.5 $, then the posterior probabilities become,
	\begin{align*}
	P(\theta = 0.2|Z = 1) =& 0.467  \\
	P(\theta = 0.1|Z = 1) =& 0.533. 
	\end{align*}
	\end{example}

\section{Bayesian Estimation and Maximum Likelihood Estimation.}
In cancer detection research, two widely used statistical approaches for estimating unknown parameters are MLE and Bayesian Estimation. MLE operates purely on the data collected during a study, aiming to find the parameter values that maximize the likelihood of observing the given outcomes. For example, when evaluating a new diagnostic test for early-stage breast cancer, MLE might be used to estimate the test’s sensitivity and specificity based on the outcomes of a trial. While effective in many cases, MLE does not incorporate any prior clinical experience or previously gathered evidence, which can be a limitation especially when dealing with rare cancers or limited sample sizes. Additionally, the confidence intervals produced through MLE rely on large sample approximations, which may be unreliable in smaller clinical studies or when events are infrequent, as often seen in early cancer detection efforts.

Bayesian Estimation, on the other hand, offers a compelling alternative in cancer diagnostics by integrating prior medical knowledge with newly observed data. For instance, if previous studies suggest that a specific biomarker is strongly predictive of prostate cancer, this information can be encoded into a prior distribution. When a new patient dataset is analyzed, the prior beliefs are updated using Bayes’ theorem to yield a posterior distribution that reflects both past and present evidence. This is particularly useful in cancer screening, where early detection relies on combining uncertain signals across multiple tests or biomarkers. Bayesian methods also produce credible intervals which are probability-based ranges for parameter estimates that can be more informative for clinicians compared to frequentist confidence intervals. As personalized oncology advances, Bayesian approaches allow researchers to tailor predictions based on individual risk factors, improving the precision and reliability of early cancer detection tools.

\begin{example}  \label{example_vaccine}
	A clinical trial is conducted to evaluate a vaccine developed for a novel strain of hepatitis. Ten individuals, all confirmed to be naturally resistant to the hepatitis virus, are selected and administered the vaccine. The immune response is assessed by measuring the level of antiserum produced in response to the vaccination. Out of the ten participants, seven exhibit a sufficient immune response. 
	
	In evaluating the effectiveness of a vaccine based on immune response data, let us define the random variable \( Z \) as the number of individuals who demonstrate an adequate response to the vaccine. We also denote \( \theta \) as the true but unknown probability that any given subject responds adequately. Since the study involves 10 independent subjects and each outcome can be classified as either a success (adequate response) or failure (inadequate response), the variable \( Z \) is naturally modeled by a Binomial distribution with parameters \( n = 10 \) and \( \theta \). That is, \( Z \overset{iid}{\sim} \text{Binomial}(10, \theta) \). Given that 7 out of the 10 individuals showed an adequate response, the likelihood of observing this outcome, conditional on a given value of \( \theta \), is expressed by the probability mass function of the Binomial distribution  
	\[
	\Pr(Z = 7 | \theta) = {10 \choose 7} \theta^7 (1 - \theta)^{(10 - 7)}.
	\]
	This likelihood function quantifies the probability of obtaining 7 responders out of 10 trials as a function of \( \theta \), and serves as the foundation for estimating the unknown parameter. To find the most likely value of \( \theta \) given the data, we turn to the method of MLE, which involves maximizing this likelihood function with respect to \( \theta \). Because working with the logarithm of the likelihood simplifies both analytical and numerical calculations especially when taking derivatives to identify critical points, we often optimize the log-likelihood instead of the original function. By applying this approach to our current problem, we find that the value of \( \theta \) that maximizes the likelihood is \( \hat{\theta}_{\text{MLE}} = 0.7 \). This estimate indicates that, based on the observed sample, the best estimate of the probability of an individual mounting an adequate immune response to the vaccine is 70\%. The maximization step can be performed either analytically using calculus or numerically through optimization routines, and the log-likelihood function plays a central role in facilitating both methods.
	
%	\Listing[store = mle,
%	caption = SAS output of Maximum likelihood estimation in Example \ref{example_vaccine}]{mle}
	
	\end{example}

When transitioning from a frequentist to a Bayesian framework for estimating the parameter \( \theta \), the probability that a subject shows an adequate response to the vaccine, we must incorporate prior information about \( \theta \), or explicitly acknowledge its absence. In cases where no reliable historical or empirical data exist to inform a prior belief, a common and reasonable choice is to adopt a non-informative or uniform prior. Specifically, one approach is to discretize the possible values of \( \theta \) and assign equal probabilities to each. For simplicity and clarity, suppose we wish to estimate \( \theta \) only up to the first decimal place. In that case, we restrict our attention to the discrete set \( \theta \in \{0.1, 0.2, 0.3, 0.4, 0.5, 0.6, 0.7, 0.8, 0.9, 1.0\} \), deliberately excluding 0 to avoid degeneracy and because a zero probability of response is biologically implausible in this context. Under this uniform prior assumption, the probability assigned to each possible value of \( \theta \) is
\[
P(\theta = 0.1) = \ldots = \Pr(\theta = 1.0) = \frac{1}{10},
\]
reflecting complete initial ignorance or neutrality about which value of \( \theta \) is more likely before observing the data.

After collecting data from the vaccine trial, in which 7 out of 10 individuals displayed an adequate response, we use Bayes' theorem to update our beliefs and derive the posterior distribution of \( \theta \). The Bayesian updating rule is given by
\[
P(\theta|Z = 7) = \frac{P(Z = 7|\theta) P(\theta)}{\Pr(Z = 7)},
\]
where the numerator consists of the likelihood of observing 7 successes given a specific value of \( \theta \), multiplied by the prior probability of that \( \theta \). The denominator, \( P(Z = 7) \), serves as a normalizing constant to ensure the resulting posterior probabilities sum to one across all considered values of \( \theta \). To evaluate this denominator, we apply the law of total probability over all ten discrete prior values
\begin{align*}
P(Z = 7) =& P(Z = 7|\theta = 0.1) \times P(\theta = 0.1) + \ldots + P(Z = 7|\theta = 1.0) \times P(\theta = 1.0) \\
=& 0.091.
\end{align*}
This final value enables computation of the posterior probabilities for each \( \theta \) by plugging in the respective likelihoods and prior probabilities into Bayes’ theorem. The resulting posterior distribution reflects how the evidence provided by the data (7 successes in 10 trials) modifies our prior beliefs, and allows us to make probabilistic inferences about the true value of \( \theta \), such as identifying the most probable value (MAP estimate) or computing credible intervals. This approach demonstrates a core strength of the Bayesian paradigm: it integrates prior knowledge (or its absence) with observed data to yield coherent and interpretable probabilistic conclusions. %The SAS codes corresponding to the posterior distribution of Table~\ref{table_posterior_theta} are provided in the Appendix.

\begin{table}[H]
	\centering
	\caption{Posterior distribution of $ \theta $ in Example \ref{example_vaccine}.}
	\label{table_posterior_theta}
	\begin{tabular}{|cc|}
		\hline
		$ \theta $ & $ P(\theta|Z = 7) $  \\
		\hline
		0.1 & 0.000  \\
		0.2 & 0.001  \\
		0.3 & 0.010  \\
		0.4 & 0.047  \\
		0.5 & 0.129  \\
		0.6 & 0.236  \\
		0.7 & 0.293  \\
		0.8 & 0.221  \\
		0.9 & 0.063  \\
		1.0 & 0.000  \\
		\hline
	\end{tabular}
\end{table}
As it has been discussed earlier, the usual posterior estimate of $ \theta $ is the posterior mean that is the mean of the posterior distribution. In this case this can be calculated as below:
\begin{align*}
\hat{\theta} =& 0.1 \times P\{(\theta|Z = 7) = 0.1\} + \ldots + 1.0 \times P\{(\theta|Z = 7) = 1.0\}  
= 0.667.
\end{align*}
Hence, to stick to the set of possible values of $ \theta $ the posterior estimate becomes 0.7. %Associated SAS code is provided in the Appendix. 
If we are tasked with determining the mode of the posterior distribution, we find that the posterior mode is 0.7, as this is the value with the highest probability mass concentrated on it. Notably, this value also corresponds exactly with the MLE, highlighting a key relationship between the two. This observation illustrates a broader and important statistical principle: when a uniform prior distribution is assumed, the posterior mode will always coincide with the MLE. This is because a uniform prior, which assigns equal probability to all values within its support, exerts no influence on the shape of the likelihood function. As a result, the posterior distribution, which is proportional to the product of the likelihood and the prior, inherits the same shape as the likelihood itself. Thus, under a uniform prior, the most probable value of the parameter according to the posterior, the mode will naturally align with the value that maximizes the likelihood function. This connection between the posterior mode and the MLE underscores the interpretability of Bayesian results in the context of classical statistical inference, especially when uninformative priors are used.

\section{Conclusion.}
Bayesian statistical approaches are redefining the landscape of clinical trial analysis by providing a framework that incorporates prior data into present studies. Unlike conventional methods that often overlook earlier findings, Bayesian techniques allow for the formal inclusion of historical insights, expert judgments, or previously collected data. This creates a more adaptive and informed model for drawing conclusions in clinical research. The process of continuously updating beliefs as fresh evidence becomes available mirrors the progressive and iterative nature of scientific inquiry in medicine. Consequently, Bayesian methods offer not just statistical rigor but also a clearer understanding of results in dynamic research settings.

The shortcomings of traditional statistics, particularly the overuse and frequent misinterpretation of p-values, have contributed to significant challenges in replicating study outcomes. The ongoing reproducibility crisis in medical science highlights the urgent need for more resilient analytical strategies. Bayesian approaches address this by enabling more nuanced evaluations of uncertainty and by naturally supporting the accumulation of evidence across studies. With the rise of efficient computational tools like MCMC, Bayesian techniques have moved from theoretical constructs to practical solutions, gaining wider acceptance within both the research community and regulatory institutions. In addition, key discrete probability models such as the Binomial, Poisson, and Negative Binomial distributions play an essential role in Bayesian analysis, especially in modeling outcomes that involve counts or binary responses. These models allow researchers to capture various patterns in clinical data, including overdispersion and event rates across time or populations. Applied correctly, they significantly improve the precision and reliability of inferences in contexts like vaccine trials or treatment effectiveness studies. Moving forward, the thoughtful integration of these statistical tools within a Bayesian framework will be crucial for enhancing the credibility and usefulness of clinical trial outcomes, ultimately benefiting patient care and public health policy.

\section*{Declarations.}
\subsection*{Ethics approval and consent to participate.}
Not applicable.
\subsection*{Consent for publication.}
Not applicable.
\subsection*{Availability of data and material.}
No data have been used.
\subsection*{Competing interests.}
The second author is an employee of Boehringer Ingelheim Pharmaceuticals. 	
\subsection*{Funding.}
Not applicable. 
\subsection*{Acknowledgements.}
Not applicable.

\section*{Appendix.}

%\begin{comment}

\subsection*{SAS codes.}

The code for Example \ref{e1}.

\begin{Sascode}[store=binom]
	DATA BINOM;
	X = PDF('BINOMIAL', 5, 0.3, 20);
	RUN;
	PROC PRINT DATA = BINOM;
	RUN;
\end{Sascode}

The code for Example \ref{e2}.

\begin{Sascode}[store=pois]
	DATA POIS;
	X = PDF('POISSON', 4, 2.1);
	RUN;
	PROC PRINT DATA = POIS;
	RUN;
\end{Sascode}

The code for Example \ref{e3}.

\begin{Sascode}[store=nbinom]
	DATA NEGBINOM;
	X = PDF('NEGBINOMIAL', 10, 0.33, 3);
	RUN;
	PROC PRINT DATA = NEGBINOM;
	RUN;
\end{Sascode}

The code for Example \ref{e4}.

\begin{Sascode}[store=cond]
	DATA COND;
	MandH = 56 + 56 + 5;
	M = 56 + 56 + 5 + 79 + 80 + 13;
	Result = MandH/M;
	RUN;
	PROC PRINT DATA = COND;
	RUN;
\end{Sascode}

The code for Example \ref{e5}.

\begin{Sascode}[store=totprob]
	DATA TOTPROB;
	S_ES = 0.8;
	ES = 0.9;
	S_LS = 0.2;
	LS = 0.1;
	S = S_ES * ES + S_LS * LS;
	RUN;
	PROC PRINT DATA = TOTPROB;
	RUN;
\end{Sascode}

The code for Example \ref{example_Bayes_theorem}.

\begin{Sascode}[store=bayes]
	DATA Bayes;
	S_ES = 0.8;
	ES = 0.9;
	S = 0.74;
	ES_S = (S_ES * ES)/S;
	RUN;
	PROC PRINT DATA = Bayes;
	RUN;
\end{Sascode}

The code for Example \ref{example_hiv}.

\begin{Sascode}[store=hiv]
	DATA HIV;
	TestPos_HIVPos = 0.95;
	HIVPos = 0.001;
	TestNeg_HIVNeg = 0.98;
	TestPos_HIVNeg = 1 - TestNeg_HIVNeg;
	HIVNeg = 1 - HIVPos;
	HIVPos_TestPos = (TestPos_HIVPos * HIVPos)/
	(TestPos_HIVPos * HIVPos + TestPos_HIVNeg * HIVNeg);
	RUN;
	PROC PRINT DATA = HIV;
	RUN;
\end{Sascode}

The code for Example \ref{example_hiv0}.

\begin{Sascode}[store = bf_hiv]
	DATA BF_HIV;
	TestPos_HIVPos = 0.95;
	TestNeg_HIVNeg = 0.98;
	TestPos_HIVNeg = 1 - TestNeg_HIVNeg;
	BF = TestPos_HIVPos/TestPos_HIVNeg;
	RUN;
	PROC PRINT DATA = BF_HIV;
	RUN;
\end{Sascode}

Additional code for Example \ref{example_hiv0}.

\begin{Sascode}[store = odds_bf_hiv]
	DATA odds_BF_HIV;
	Prior_Odds = 0.001/(1 - 0.001);
	BF = 47.5;
	Posterior_odds = BF * Prior_Odds;
	RUN;
	PROC PRINT DATA = odds_BF_HIV;
	RUN;
\end{Sascode}

Code for figure \ref{fig:1}.

\begin{Sascode}[store = odds_bf_hiv]
	data hiv_probabilities;
	length Hypothesis $20 ProbabilityType $10;
	input Hypothesis $ ProbabilityType $ Probability;
	datalines;
	HIV_Positive Prior 0.001
	HIV_Negative Prior 0.999
	HIV_Positive Posterior 0.045
	HIV_Negative Posterior 0.955
	;
	run;
	
	proc sgplot data=hiv_probabilities;
	title "Updating Beliefs with Bayes Factor: Prior vs 
	Posterior Probabilities of HIV Status";
	vbarparm category=Hypothesis response=Probability / 
	group=ProbabilityType groupdisplay=cluster datalabel;
	yaxis label="Probability" values=(0 to 1 by 0.1);
	xaxis label="Hypothesis";
	keylegend / title="Probability Type";
	run;
	\end{Sascode}	

R code for Figures \ref{fig:2} and \ref{fig:3}.
\begin{Sascode}
library(igraph)
library(bnlearn)
num_nodes <- 500
node_names <- paste0("V", 1:num_nodes)
set.seed(42)
dag_mat <- matrix(0, nrow = num_nodes, ncol = num_nodes)
colnames(dag_mat) <- rownames(dag_mat) <- node_names
edge_prob <- 0.01  # Sparsity control
for (i in 1:(num_nodes - 1)) {
for (j in (i + 1):num_nodes) {
if (runif(1) < edge_prob) {dag_mat[i, j] <- 1}}}  
edge_list <- which(dag_mat == 1, arr.ind = TRUE)
edge_df <- data.frame(
from = rownames(dag_mat)[edge_list[, 1]],
to   = colnames(dag_mat)[edge_list[, 2]])
dag_igraph <- graph_from_data_frame(edge_df, vertices = node_names, directed = TRUE)
plot(dag_igraph,
layout = layout_with_sugiyama(dag_igraph)\$layout,
vertex.color = "darkgreen",
vertex.label.color="white",edge.width=1,edge.color="black",
vertex.label.cex = 0.6,vertex.size = 12,edge.arrow.size = 0.4)
\end{Sascode}

Code for marginal likelihood in Example~\ref{example_coin_toss}.
	\begin{Sascode}[store = cointoss]
		DATA COINTOSS;
		INPUT THETA PRIOR;
		DATALINES;
		0.35 0.5
		0.5  0.5
		;
		RUN;
		DATA COINTOSS;
		SET COINTOSS;
		LIKELIHOOD = PDF('BINOMIAL', 6, THETA, 10) * 
		PDF('BINOMIAL', 4, THETA, 7);
		LIKELIHOODPRIOR = LIKELIHOOD * PRIOR;
		RUN;
		PROC MEANS DATA = COINTOSS SUM;
		VAR LIKELIHOODPRIOR;
		RUN;
	\end{Sascode} 

Code for posterior distribution in Example~\ref{example_coin_toss}.
	\begin{Sascode}[store = post_cointoss]
		DATA COINTOSS;
		SET COINTOSS;
		POSTERIOR = LIKELIHOODPRIOR/0.0330077;
		RUN;
		PROC PRINT DATA = COINTOSS;
		RUN;
	\end{Sascode}

Code for Bayes factor in Example~\ref{example_coin_toss}.
\begin{Sascode}[store = BFcointoss]
	DATA BF_COINTOSS;
	BF = 0.056076/0.0099389;
	RUN;
	PROC PRINT DATA = BF_COINTOSS;
	RUN;
\end{Sascode}

Code for Example~\ref{example_blood}.
\begin{Sascode}[store = blood]
	DATA BLOOD;
	INPUT THETA PRIOR;
	DATALINES;
	0.2 0.25
	0.1 0.75
	;
	RUN;
	DATA BLOOD;
	SET BLOOD;
	LIKELIHOOD = PDF('NEGBINOMIAL', 7, theta, 1);
	LIKELIHOODPRIOR = LIKELIHOOD * PRIOR;
	RUN;
	PROC MEANS DATA = BLOOD SUM;
	VAR LIKELIHOODPRIOR;
	RUN;
	DATA BLOOD;
	SET BLOOD;
	POSTERIOR = LIKELIHOODPRIOR/0.0463580;
	RUN;
	PROC PRINT DATA = BLOOD;
	RUN;
\end{Sascode}

Code for Example~\ref{example_blood} with uniform prior.
\begin{Sascode}[store = bloodunif]
	DATA BLOODUNIF;
	INPUT THETA PRIOR;
	DATALINES;
	0.2 0.5
	0.1 0.5
	;
	RUN;
	DATA BLOODUNIF;
	SET BLOODUNIF;
	LIKELIHOOD = PDF('NEGBINOMIAL', 7, theta, 1);
	LIKELIHOODPRIOR = LIKELIHOOD * PRIOR;
	RUN;
	PROC MEANS DATA = BLOODUNIF SUM;
	VAR LIKELIHOODPRIOR;
	RUN;
	DATA BLOODUNIF;
	SET BLOODUNIF;
	POSTERIOR = LIKELIHOODPRIOR/0.0448864;
	RUN;
	PROC PRINT DATA = BLOODUNIF;
	RUN;
\end{Sascode}

Code for Example~\ref{example_vaccine}.
\begin{Sascode}[store = mle]
PROC NLP;
MAX loglikelihood;
PARMS THETA = 0.5;
loglikelihood = log(PDF('BINOMIAL', 7, THETA, 10));
RUN;
\end{Sascode}

Code for Table~\ref{table_posterior_theta}.
\begin{Sascode}[store = vaccine]
	DATA VACCINE;
	INPUT THETA PRIOR;
	DATALINES;
	0.1 0.1
	0.2 0.1
	0.3 0.1
	0.4 0.1
	0.5 0.1
	0.6 0.1
	0.7 0.1
	0.8 0.1
	0.9 0.1
	1.0 0.1
	;
	RUN;
	DATA VACCINE;
	SET VACCINE;
	LIKELIHOOD = PDF('BINOMIAL', 7, theta, 10);
	LIKELIHOODPRIOR = LIKELIHOOD * PRIOR;
	RUN;
	PROC MEANS DATA = VACCINE SUM;
	VAR LIKELIHOODPRIOR;
	RUN;
	DATA VACCINE;
	SET VACCINE;
	POSTERIOR = LIKELIHOODPRIOR/0.0909993;
	RUN;
	PROC PRINT DATA = VACCINE;
	RUN;
\end{Sascode}

Code for Posterior value 0.7.
\begin{Sascode}[store = posteriormean]
	DATA VACCINE;
	SET VACCINE;
	EXPECTATION = THETA * POSTERIOR;
	RUN;
	PROC MEANS DATA = VACCINE SUM;
	VAR EXPECTATION;
	RUN;
\end{Sascode}

R code for Figure~\ref{fig:4}.
\begin{Sascode}
TestResult_levels <- c("Positive", "Negative")
Risk_levels <- c("LowRisk", "HighRisk")
imaginary_samples <- c(1, 2, 5, 10, 20, 50, 100, 200, 500)
Risk_vec <- c()
n_vec <- c()
Result_vec <- c()
Prob_vec <- c()
converge <- function(n, start, end, rate = 50) {
	return(end + (start - end) * exp(-n / rate))
}
for (n in imaginary_samples) {
	prob_high_positive <- converge(n, start = 0.9, end = 0.5) 
	prob_low_positive  <- converge(n, start = 0.1, end = 0.5) 
	prob_high <- c(Positive = prob_high_positive, Negative = 1 - prob_high_positive)
	prob_low  <- c(Positive = prob_low_positive,  Negative = 1 - prob_low_positive)
	for (Risk in Risk_levels) {
		probs <- if (Risk == "LowRisk") prob_low else prob_high
		for (i in 1:2) {
			Risk_vec   <- c(Risk_vec, Risk)
			n_vec      <- c(n_vec, n)
			Result_vec <- c(Result_vec, names(probs)[i])
			Prob_vec   <- c(Prob_vec, probs[i])}}}
results <- data.frame(Risk = Risk_vec, n = n_vec, TestResult = Result_vec,\\
Prob = Prob_vec)
results_low <- results[results$Risk == "LowRisk", ]
results_high <- results[results$Risk == "HighRisk", ]
pdf('C:/Users/Paramahansa Pramanik/OneDrive/Desktop/PAPERS_ME/Arnab_book\\
/Manuscript/file4.pdf')
par(mfrow = c(2, 1), mar = c(4, 4, 2, 1))
plot(results_high$n[results_high$TestResult == "Positive"], 
results_high$Prob[results_high$TestResult == "Positive"],
type = "l", log = "x", col = "blue", lty = 3, ylim = c(0, 1),
xlab = "Total number of patients (log scale)", 
ylab = "P(TestResult | HighRisk)",
main = "Estimated HIV Test Result Probabilities (High Risk)")
lines(results_high$n[results_high$TestResult == "Negative"], 
results_high$Prob[results_high$TestResult == "Negative"],
col = "red", lty = 4)
legend("topright", legend = c("Positive ", "Negative "), 
col = c("blue", "red"), lty = c(3, 4))
plot(results_low$n[results_low$TestResult == "Positive"], 
results_low$Prob[results_low$TestResult == "Positive"],
type = "l", log = "x", col = "blue", lty = 3, ylim = c(0, 1),
xlab = "Total number of patients (log scale)", 
ylab = "P(TestResult | LowRisk)",
main = "Estimated HIV Test Result Probabilities (Low Risk)")
lines(results_low$n[results_low$TestResult == "Negative"], 
results_low$Prob[results_low$TestResult == "Negative"],
col = "red", lty = 4)
legend("topright", legend = c("Positive", "Negative"), 
col = c("blue", "red"), lty = c(3, 4))
dev.off()
	
\end{Sascode}	

%\end{comment}

%\bibliographystyle{plainnat}
%\bibliographystyle{apalike}
\bibliographystyle{abbrv}
\bibliography{bib}

\end{document}